\documentclass[a4paper, notoc]{article}
\usepackage{jcappub}
\usepackage{amsfonts}
\usepackage{wasysym}
\usepackage{amssymb}
\usepackage{epsfig}
\usepackage{textcomp}
\usepackage{graphicx,psfrag}
\usepackage[caption=false]{subfig}
\usepackage{braket}
\usepackage{placeins}%\usepackage{ifpdf}

 \newlength{\wth}
\newcommand{\be}{\begin{equation}}
\newcommand{\ee}{\end{equation}}
\newcommand{\bea}{\begin{eqnarray}}
\newcommand{\eea}{\end{eqnarray}}
\newcommand{\nn}{\nonumber}

\newcommand{\ti}{\times}
\newcommand{\half}{\frac{1}{2}}
\newcommand{\mc}{\mathcal}

\newcommand{\beqa}{\begin{eqnarray}}
\newcommand{\eeqa}{\end{eqnarray}}

\newcommand{\kpc}{\hbox{\,kpc}}
\newcommand{\Mpc}{\hbox{\,Mpc}}
\newcommand{\keV}{\hbox{\,keV}}
\newcommand{\beq}{\begin{equation}}
\newcommand{\eeq}{\end{equation}}
\newcommand{\mug}{\,\mu\hbox{G}}
\newcommand{\parphoton}{\Ket{\gamma_{\parallel}}}
\newcommand{\perpphoton}{\Ket{\gamma_{\perp}}}
\newcommand{\axion}{\Ket{a}}
\newcommand{\dd}{\hbox{d}}
\newcommand{\maxi}{\rm max}
\newcommand{\mini}{\rm min}

\title{Soft X-ray Excess in the Coma Cluster from a Cosmic Axion Background}

\author[a]{Stephen Angus,}
\author[a]{Joseph P. Conlon,}
\author[a]{M.C.~David Marsh,}
\author[a]{Andrew J.~Powell}
\author[a,b]{and Lukas T.~Witkowski}
\affiliation[a]{Rudolf Peierls Centre for Theoretical Physics, University of Oxford,\\ 1 Keble Road, Oxford OX1 3NP, United Kingdom}
\affiliation[b]{Institut f\"ur Theoretische Physik, Universit\"at Heidelberg, \\ Philosophenweg 19, 69120 Heidelberg, Germany}

\emailAdd{stephen.angus@physics.ox.ac.uk}
\emailAdd{j.conlon1@physics.ox.ac.uk}
\emailAdd{david.marsh1@physics.ox.ac.uk}
\emailAdd{andrew.powell2@physics.ox.ac.uk}
\emailAdd{l.witkowski@thphys.uni-heidelberg.de}

\abstract{ We show that the soft X-ray excess in the Coma cluster can be explained by a cosmic background of relativistic axions converting into photons in the cluster magnetic field. We provide a detailed self-contained review of the cluster soft X-ray excess, the proposed astrophysical explanations and the problems they face, and explain how a $0.1-1\keV$ axion background naturally arises at reheating in many string theory models of the early universe. We study the morphology of the soft excess by numerically propagating axions through stochastic, multi-scale magnetic field models that are consistent with observations of Faraday rotation measures from Coma. By comparing to ROSAT observations of the $0.2 - 0.4\keV$ soft excess, we find that the overall excess luminosity is easily reproduced for $g_{a\gamma\gamma} \sim 2 \ti 10^{-13}\,$GeV$^{-1}$. The resulting morphology is highly sensitive to the magnetic field power spectrum. For Gaussian magnetic field models, the observed soft excess morphology prefers magnetic field spectra with most power in coherence lengths 
on ${\cal O}(3\kpc)$ scales over those with most power on ${\cal O}(12\kpc)$ scales. Within this scenario, we bound the mean energy of the axion background to $50\,{\rm eV}\lesssim \langle E_a \rangle \lesssim 250\,{\rm eV}$, the axion mass to  $m_a \lesssim 10^{-12}\,\hbox{eV}$, and derive a lower bound on the axion-photon coupling $g_{a\gamma\gamma} \gtrsim \sqrt{0.5/\Delta N_{\rm eff}}\, 1.4 \ti 10^{-13}\,$GeV$^{-1}$.
}

\begin{document}
\maketitle

\section{Introduction}

This paper
 provides the first systematic study of
a proposed physical phenomenon
which would forge
a linkage between three disparate topics: an excess in the spectrum of soft X-rays from galaxy clusters,
dark radiation, and string theory models of the early universe. %In brief,
With the Coma cluster as our prime example we give  a detailed demonstration of how
the observed cluster soft X-ray excess may emerge from axion-photon conversion of a
homogeneous dark radiation $0.1 - 1\keV$ Cosmic Axion Background (CAB), which in turn arose from moduli decays in the early universe
\cite{CAB}.

Galaxy clusters are the largest gravitationally bound objects in the universe and have historically served as powerful indicators of novel fundamental physics \cite{Zwicky}.
  In addition to the dark matter component comprising around 80\% of the cluster mass,
  around 15\% of the mass is in a hot ionised intracluster medium (ICM) with typical temperatures
 of $T \approx 10^8\, $K (corresponding to $\omega \approx 7\keV$) and number densities $n \sim 10^{-1}-10^{-3}~\hbox{\,cm}^{-3}$.
 The ICM represents the large majority of a cluster's baryonic mass and generates diffuse X-ray emission through thermal bremsstrahlung.

A thermal bremsstrahlung spectrum gives to a good approximation  a constant emissivity per unit energy at low energies. However,
 observations of a large number of galaxy clusters have found evidence at low energies around $E \lesssim 0.4\keV$, for excess emission
 above that from the hot ICM. This soft excess was initially observed in the
 Virgo and  Coma clusters in 1996 \cite{Lieu1996a, Lieu1996b, Bowyer96} and has since been found in many other clusters \cite{astroph0205473, 08010977}.
There are two candidate astrophysical explanations: emission from a warm $T \approx 0.1$ keV gas; and inverse-Compton scattering of $\gamma \sim 300-600$ non-thermal electrons
on the Cosmic Microwave Background (CMB). The former explanation has difficulty with rapid cooling times of a warm gas and the lack of associated ${\rm O}_{\rm VII}$ line emission; the latter
has difficulty remaining consistent with the observed level of synchrotron radio emission and the failure to detect clusters in gamma rays.  In section
\ref{sec:excess} we provide a detailed review of the soft excess phenomenon and these proposed astrophysical explanations.

Recently, in \cite{CAB},  the cluster soft excess was proposed to arise from conversion of a primordial Cosmic Axion Background into photons in the magnetic field of galaxy clusters.
As we will review in section 3, the existence of such a  background of
highly relativistic axions (or axion-like particles) is theoretically well-motivated in  models of the early universe arising from compactifications of
string theory to four dimensions. The axions arise from moduli decays at the time of reheating and would
linger today as a homogeneous and isotropic  Cosmic Axion Background with a non-thermal spectrum determined by the expansion of the universe during the time of moduli decay, as discussed in \cite{13041804}. For moduli masses $m_{\Phi} \approx 10^6\hbox{\,GeV}$ the present energy of these axions is
 $E_a \sim 0.1-1\keV$.

The existence of such a CAB can be indirectly probed through its contribution to dark radiation.
Dark radiation is traditionally parametrised in terms of an effective number of neutrino species, $N_\mathrm{eff} = 3.046 + \Delta N_\mathrm{eff}$, where the first term corresponds to
three neutrino species undergoing thermal decoupling. The relativistic contribution to the energy density at
CMB decoupling can be written as
\be
\rho_\mathrm{radiation} = \rho_{\gamma} \left( 1 + \left( \frac{7}{8} \right) \left( \frac{4}{11} \right)^{4/3} N_\mathrm{eff} \right).
\ee
The amount of dark radiation can be probed either by measurements of the CMB or through measurements of primordial BBN abundances.
There are current hints at the $1$-$3$ sigma level for a non-zero value of $\Delta N_{eff}$.
Recent results from
the Planck satellite give $N_\mathrm{eff} = 3.52 \pm 0.24$ (CMB + BAO + H$_0$) or $N_\mathrm{eff} = 3.30 \pm 0.27$ (CMB + BAO) \cite{13035076}, depending on whether measurements of the Hubble constant in the local universe are included in the combination.
Recent BBN-only measurements based on primordial Helium and Deuterium abundances give $N_\mathrm{eff} = 3.50 \pm 0.20$ \cite{PettiniCooke}.

In the presence of a magnetic field, axions can  directly convert into photons via the coupling,
\be
{\cal L}_{a\gamma\gamma} = \frac{1}{8M} a F_{\mu \nu} \tilde F^{\mu \nu} = \frac{1}{M} a {\bf E} \cdot {\bf B}   \equiv g_{a \gamma \gamma} a {\bf E} \cdot {\bf B} \, .
\ee
In the enlightening case of sufficiently high axion energies or small ambient electron densities, the conversion probability for a fixed domain is given by
\be
P_{a\rightarrow \gamma} = \frac{1}{4} \left( \frac{B_{\perp} L}{M}\right)^2 \, ,
\ee
where $B_{\perp}$ denotes the magnetic field component transverse to the axion velocity and $L$ denotes the corresponding coherence length \cite{Sikivie:1983}.
This conversion allows the potential detection of a CAB through axion-photon conversion.

Galaxy clusters support magnetic fields that are modest in magnitude ($B \approx \mu$G) but are extended over megaparsec distances and have kiloparsec coherence scales, allowing
observationally significant axion-photon conversion probabilities.
In \cite{CAB}, a crude single-domain model with a fixed magnitude and coherence length for the magnetic field was used to estimate the axion-photon coupling $M$ that would be required to reproduce the soft excess in Coma from a CAB, finding $M \approx 10^{13}\,  \hbox{GeV}$.

In this paper we continue the study of axion-photon conversion in the Coma cluster using a far more detailed model of the Coma magnetic field.
This model was constructed in \cite{10020594} to fit
rotation measure (RM) observations of seven polarised light sources, using the Coma cluster as a Faraday screen.
The model describes the central $\hbox{Mpc}^3$ of Coma (see also \cite{13057228} for a magnetic field model describing the region 1.5 Mpc to the southwest of the cluster centre).
We review the observational evidence for cluster magnetic fields in section \ref{sec:B} and describe the  model for the Coma magnetic field in detail in section \ref{sec:ComaB}.
Using this stochastic model, we construct a numerical simulation of the magnetic field in the central region of the cluster, propagate axions through it and quantitatively study the resulting predictions for the soft excess morphology.

This paper is organised as follows.
Section \ref{sec:excess} reviews the cluster soft excess phenomenon, the proposed astrophysical explanations and the constraints
on these explanations. Section \ref{sec:CAB} describes how a Cosmic Axion Background arises naturally in string models of the early universe. Section \ref{sec:B} reviews the magnetic field model proposed in \cite{10020594} for the Coma cluster. This volume of review is larger than standard, %subliminal message to make people think of LVS? --SA :-) JC
but necessary to make the paper
self-contained given the disparate topics involved. In section \ref{sec:analysis} we describe the results of simulated axion-photon conversion and
compare the simulations to observations
of the soft excess.  In section \ref{sec:concl} we conclude.

%%%%%%%%%%%%%%%%%%%%%%%%%%%%%%%%%%%%%%%%%%%%%%%%%%%%%%%%%%%
%%%%%%%%%%%%%%%%%%%%%%%%%%%%%%%%%%%%%%%%%%%%%%%%%%%%%%%%%%%
%%%%%%%%%%%%%%%%%%%%%%%%%%%%%%%%%%%%%%%%%%%%%%%%%%%%%%%%%%%

\section{Review of soft excess observations}\label{sec:other} \label{sec:excess}

The aim of this section is to summarise the observational history of the soft excess from galaxy clusters and to discuss the astrophysical models proposed to account for it.
We also review the status of these models in light of more recent
 measurements of the Coma magnetic field as well as the (null)
observations of galaxy clusters in gamma rays.
A complementary review of the soft excess phenomenon from 2008 is given in \cite{08010977}.

%%%%%%%%%%%%%%%%%%%%%%%%%%%%%%%%%%%%%%%%%%%%%%%%%%%%%%%%%%%
%%%%%%%%%%%%%%%%%%%%%%%%%%%%%%%%%%%%%%%%%%%%%%%%%%%%%%%%%%%

\subsection{Preliminaries: X-ray observations of galaxy clusters}

We first review the most basic aspects of observing X-rays from galaxy clusters.
This material will be familiar and indeed elementary to astrophysicists,
but is not part of the standard particle theory education, and we include
it for the sake of completeness.

Galaxy clusters are the largest virialised structures in the universe, with typical masses $\sim 10^{14}-10^{15} M_{\astrosun}$ and spatial extents of $\mc{O}(1\hbox{\,Mpc})$. Clusters emit light of all frequencies ranging from radio waves to X-rays. The space between galaxies is suffused with an energetic ionised plasma, termed the intracluster medium (ICM).
The energy of the ICM arose from the release of
gravitational potential energy as the cluster formed through accretion and merger of subclusters.
 The hot ICM has a temperature $T \sim 2-8\,\hbox{keV}$ and emits X-ray emission via thermal bremsstrahlung,
 with typical X-ray luminosities of $\mc{L} \sim 10^{42 - 45}\hbox{\,erg s}^{-1}$ (where $1.6~\!\ti~\!10^{-3}\hbox{\,erg} = 1\hbox{\,GeV}$).
  The energy flux from thermal bremsstrahlung is
\cite{Longair}
\bea
I(\nu) & = & \frac{1}{3 \pi^2} \left( \frac{\pi}{6} \right)^{\half} \frac{Z^2 e^6}{\epsilon_0^3 c^3 m_e^2} \left( \frac{m_e}{kT} \right)^{\half}
g(\nu, T) N N_e \exp \left( - \frac{h \nu}{kT} \right) \nn \\
& = & A Z^2 N N_e  \frac{ g(\nu, T)  \exp \left(-\frac{h \nu}{kT} \right)}{\sqrt{kT}},
\eea
where in the second equation $A$ is a constant and we have restricted to the key parameters. Here $N$ is the ion density, $N_e$ the electron density, and $Z$ the ionic charge.
The Gaunt factor $g(\nu, T)$ is a slowly varying function of frequency, and so at photon energies much less than the temperature
 the emitted flux is approximately constant as a function of frequency (note that this implies a photon number index of $-1$, as $\dd N_{\gamma}(\nu)/\dd E \sim \nu^{-1}$).
The low-frequency emissivity then scales quadratically with density and inversely with the square-root of the temperature. This implies that thermal emissivity within a fixed waveband is minimised for high temperatures and low densities.\footnote{As will become relevant in section \ref{sec:analysis}, we note that both these conditions are satisfied in the Coma cluster.}

Extragalactic X-rays must reach earth by passing through the Milky Way where they may  be attenuated by absorption.
The extent of the absorption is determined by the effective neutral Hydrogen (NH) column density.
Here, `effective' refers to the fact that helium also contributes significantly to absorption.
Within the galaxy this column density varies from a global minimum of $\approx 5 \ti 10^{19}\hbox{\,cm}^{-2}$ at the Lockman hole
to around $ 1 \ti 10^{22}\hbox{\,cm}^{-2}$
towards the galactic centre. Plots of absorption fractions as a function of frequency can be found in e.g. \cite{Seward}.

The soft excess frequencies considered here are in the extreme ultraviolet/soft \hbox{X-ray} bands, with an approximate range of $0.1 - 1\keV$. Light at such frequencies is heavily absorbed: for \hbox{$N_H = 1 \ti 10^{20}\hbox{\,cm}^{-2}$}, a $200\,$eV photon has a $\approx45 \%$ transmission probability, while for $N_H > 10^{21}\hbox{\,cm}^{-2}$ the transmission probability is effectively zero. In comparison, photons with energies in the $1-2\keV$ range have transmission probabilities ranging from $8\% - 60\%$ even for $N_H = 10^{22}\hbox{\,cm}^{-2}$.  Thus the galactic plane is opaque
to extragalactic soft X-rays,
but at high galactic latitude ($\vert b \vert \gtrsim 30^{\circ}$) absorption is sufficiently limited that an extragalactic flux can be measured,
and clusters and other extragalactic objects can be observed.

It follows that an accurate measurement of extragalactic soft photon fluxes requires an accurate measurement of the column densities.
These are measured using all-sky 21cm surveys, such as Dickey and Lockman (1990) \cite{DickeyLockman}  or the Leiden/Argentine/Bonn (LAB) survey (2005) \cite{Kalberla}. For certain soft
excess measurements as e.g. the sample of 38 clusters in \cite{astroph0205473}, high resolution 21cm observations from the Green Bank NRAO have also been used to ensure uniformity of the NH column density on the scale of a cluster.

Accurate measurements of soft X-ray fluxes from distant sources also require accurate measurements of the local soft X-ray background. The soft X-ray sky has gradients on the scale of a few degrees and is also subject to temporal variation based on solar flares and charge exchange scattering between the solar wind and the Earth's exosphere. Galaxy clusters are large objects, with a typical diameter $d \approx 1\hbox{\,Mpc}$, and the intracluster gas generates diffuse emission with relatively low surface brightness.
The ideal background is therefore one which is both spatially and temporally contiguous. This is most easily accomplished if the observing telescope has a large field of view, which can then accommodate the entire cluster and allow the background to be taken from the edge region of the detector.

%%%%%%%%%%%%%%%%%%%%%%%%%%%%%%%%%%%%%%%%%%%%%%%%%%%%%%%%%%%
%%%%%%%%%%%%%%%%%%%%%%%%%%%%%%%%%%%%%%%%%%%%%%%%%%%%%%%%%%%

\subsection{History of soft excess observations}

Soft X-ray emission from galaxy clusters has been consistently found by a number of satellites and in a significant number of clusters. In this section we review how the different satellite missions have contributed to observational evidence for the excess.
\subsubsection{The EUVE discovery}
The original discovery of the cluster soft excess phenomenon was made in 1996 %by Lieu et al
with observations of the Virgo and Coma clusters \cite{Lieu1996a, Lieu1996b, Bowyer96} using the Extreme Ultraviolet Explorer (EUVE).
The EUVE Lex/B detector \cite{BowyerEUV} had peak response at $138\hbox{\,eV}$ and
a passband of $65-248\hbox{\,eV}$ at 10\% of peak, although with very limited spectral resolution.
The %EUVE had a large rectangular
 field of view was large and rectangular, covering $2^{\circ} \ti 40^{'}$.
 The properties of the ICM gas were determined by measurements on complementary instruments with X-ray sensitivity.
The EUVE observations revealed a
large EUV excess over the level of emission expected from thermal bremsstrahlung from the hot gas.
By default, %analogue with the hot gas,
these observations were interpreted as
evidence for a warm gas component ($T \approx 5 \ti 10^5\,\mathrm{K} \simeq 50\hbox{\,eV}$) within the cluster.
These observations of Virgo and Coma were then extended to the Abell clusters A1795 and A2199 \cite{Mittaz98, astroph9902300}, where very large
soft excesses, of up to 600\% of the hot gas emission, were reported.

Disagreement then arose about these observations, concerning in particular the details of background subtraction for the EUVE satellite
and the variation in telescope sensitivity over the field of view. The EUVE data for various clusters was reanalysed in
%by Bowyer, Berghofer and Korpela
\cite{astroph9911001, astroph9912421, astroph0011274, astroph0101300},
which found no soft excess for Fornax, A1795, A2199 and A4059. These studies did however confirm the existence of a soft excess in Coma and Virgo, although disagreeing with the original works on the detailed properties. However these analyses were themselves again
challenged, with re-observations
of these clusters with \emph{in situ} background measurements reconfirming the presence of an excess \cite{astroph9910298, astroph0011186}. A further reanalysis of the EUVE and ROSAT observations for these clusters was performed using wavelet techniques \cite{astroph0204345}. This analysis
found that the EUV and X-ray populations were statistically different and found
EUV soft excesses present in Coma, Virgo, A1795, A2199 and A4059. The magnitude of the soft excess became more pronounced at large radii.

These cluster observations represent the limit of EUVE: the satellite was
decommissioned in January 2001 and re-entered the atmosphere in 2002.
While all analyses were in agreement over the existence of a soft excess in the Coma and Virgo clusters,
no consensus was reached over the existence of an EUV soft excess in the other clusters observed.

%%%%%%%%%%%%%%%%%%%%%%%%%%%%%%%%%%%%%%%%%%%%%%%%%%%%%%%%%%%

\subsubsection{Consolidation by ROSAT}
The other main satellite in which the soft excess has been observed was the ROSAT satellite.
ROSAT \cite{ROSAT} was a joint German-UK-US mission, operating from 1990 to 1999, which carried out an all-sky survey in the $0.1-2.4\keV$ waveband.
ROSAT has been used to study the soft excess both through individual papers on particular clusters and through a large statistical survey of 38 clusters by Bonamente et al \cite{astroph0205473}. We first discuss the individual papers before describing the statistical sample of
\cite{astroph0205473}.

In many ways ROSAT was an ideal
satellite for studying the soft excess: in particular, it was more suitable for this purpose than the next-generation satellites Suzaku, Chandra and XMM-Newton.
It had an energy range of $0.1-2.4\keV$ with a large field of view of $2^{\circ}$ diameter and an effective area
of $200\hbox{\,cm}^2$ at $0.28\keV$.
 The PSPC detectors had a low and stable internal background and a well-calibrated effective area.
The spectral resolution was however limited, with $\Delta E/E = 0.43 \sqrt{0.93\keV/E}$.
The former properties are ideal for the study of large-scale diffuse emission from clusters, and
the low spectral resolution is not a fatal disadvantage.
 The large field of view enables a single pointing to encompass the whole cluster, allowing a contiguous background from the
outskirts of the image. The low internal background, while less relevant for bright point sources, is crucial for accurate measurements of weak diffuse emission.

Four clusters within the Shapley supercluster --- A3558, A3560, A3562 and A3571  --- were studied in \cite{astroph0103331} using a combination of ROSAT and Bepposax data. ROSAT data revealed a soft excess in all clusters, with the much less sensitive Bepposax also finding
the excess in A3571.
A study of ROSAT observations of clusters in the Hercules supercluster \cite{astroph0512591} also found soft excesses, although background subtraction in this case is particularly difficult as these clusters lie in the direction of the North Polar Spur, a large soft X-ray emitter. In \cite{astroph0107379} very strong soft excess emission was reported for Sersic-159 (a.k.a.~AS1101). However, it was subsequently determined that the strength of this excess was an artefact based on the use of an artificially high NH value $(1.8 \ti 10^{20}\hbox{\,cm}^{-2})$ from the Dickey and Lockman survey, which was revised down to $1.15 \ti 10^{20}\hbox{\,cm}^{-2}$ in the LAB survey. While the soft excess remains, it is
now at a level comparable to other clusters \cite{11070932}.

The centre of the Coma cluster was studied in \cite{astroph0205473}, and two studies have also been performed for the
outskirts of the cluster. In the centre of the cluster, a large soft excess was found at very high statistical significance. We will
describe the magnitude and morphology of the central excess in section 5 when we compare to our simulations of axion-photon conversion.
A study of cluster outskirts, where there is less intrinsic signal, requires a careful subtraction of the background. The first study (2002) \cite{astroph0211439} used offset pointings to measure the background and found a soft excess extending out to
$2.6\hbox{\,Mpc}$. A later study (2009) was \cite{09033067}, which used the observation of Coma during the ROSAT all-sky survey. Although this observation was of limited temporal duration, it involves a background that is both spatially and temporally contiguous, being measured as the satellite slews across the sky. This found a soft excess in the R2 channel ($0.14-0.28\keV$) extending up to $5\hbox{\,Mpc}$ from the cluster centre.

The largest study to date of the soft excess was carried out in \cite{astroph0205473}, in which
%This was a survey of
38 clusters with favourable observational conditions were studied. %for studying the soft excess.
The requirements for inclusion were that a cluster lay
at high galactic latitude and low column density, and that both a pointed ROSAT observation and narrow-beam column density measurements from the Green Bank NRAO existed.
Note that for these clusters there have been no large revisions of the column densities between the values given in \cite{astroph0205473} and the LAB values.
This study looked for excess emission in the ROSAT $0.2 - 0.4\keV$ band compared to expectations from the hot gas.
In most cases (except for the nearest clusters), the large ROSAT field of view enabled an
\emph{in situ} background measurement from the peripheral detector regions.

The study found that soft excess emission was a general feature of galaxy clusters.
A statistically significant detection was observed in $~30\%$ of the sample|in some cases, such as Coma, at very high statistical significance. Soft excess was only ruled out in a small number of clusters using the ROSAT data.
The clusters studied covered a range in redshift from 0.0043 (Virgo) to 0.308 (A2744), and
soft excess emission was found for both nearby and distant clusters, for both relaxed and
disturbed clusters, and at a wide range of column densities $N_H = 1 $--$ 4.5 \ti 10^{20}\hbox{\,cm}^{-2}$.
In all cases the relative magnitude of the soft excess compared to the thermal ICM emission did not exceed 30 percent.
The soft excess appeared to follow an
identifiable morphological trend in that %is that
excess emission was preferentially found at radii $r \gtrsim 175\kpc$, outside the very centre of clusters.
In section \ref{sec:analysis} we will comment on how the properties of axion-photon conversion in cluster magnetic fields
may explain this trend.
%%%%%%%%%%%%%%%%%%%%%%%%%%%%%%%%%%%%%%%%%%%%%%%%%%%%%%%%%%%

\subsubsection{The new generation satellites: XMM-Newton, Suzaku and Chandra}

Soft excess emission has also been studied with the XMM-Newton satellite. This offers a much greater spectral resolution than ROSAT
(roughly 60 eV at 0.5 keV) and an energy range that extends to the hot gas temperatures, but at the cost of a much smaller field of view ($30^{'}$ diameter as compared to $114^{'}$) and a larger detector background. The small field of view makes it harder to perform an accurate background subtraction.
Statistical studies of large samples of clusters with XMM-Newton were carried out in \cite{astroph0210610, astroph0210684, astroph0305424, KaastraDec04}, and ${\rm O}_{\rm VII}$ line emission (which would have confirmed a thermal origin of the soft excess) was reported in \cite{astroph0210684, astroph0305424, astroph0309019}. However, it was subsequently determined that these lines could arise through the local or galactic
background \cite{astroph0501007, astroph0602527, astroph0610461}, while further direct searches for emission lines proved null
\cite{astroph0608072, 07040475, 10103840, 11104422} (although see \cite{astroph0610424} for a marginal detection).

The XMM-Newton detections of soft excess were challenged in \cite{astroph0602527}, where it was argued that they arose from an incorrect background
subtraction. This issue was therefore revisited in \cite{astroph0610461}. Here it was found that while the background for the bright cluster-centre regions was not large
enough for background subtraction to be relevant, the recommended calibration for the two EPIC instruments (MOS and PN) had changed since the 2003 analyses. With the new calibrations, these authors found that soft excesses were consistently detected by
 one of the two EPIC instruments (MOS) while being absent in the other (PN). For this reason the current state of soft excess detection in XMM-Newton
 remains unclear.

Let us also mention the (limited) studies of the soft excess that have been carried out with Suzaku and Chandra. The limited field of view
again makes these suboptimal instruments for studies of diffuse soft emission.
The soft excess emission for the cluster A3112 was studied in \cite{07070992} (with Chandra) and in \cite{10103840} (with Suzaku).
Earlier studies had suggested a very strong soft excess in this cluster, but
A3112 is a cluster for which the revised LAB absorption column value ($1.3 \ti 10^{20}\hbox{\,cm}^{-2}$) was significantly lower than the Dickey and Lockman value ($2.6 \ti 10^{20}\hbox{\,cm}^{-2}$). The latter study, using the revised LAB value, found the soft excess to still be present but (as expected) reduced compared to the earlier analyses.

A similar case is the cluster Sersic 159 (AS1101), studied with Suzaku in \cite{07040475}. This study confirmed the existence of the soft excess but did not find the tentative ${ \rm O}_{\rm VII}$ emission line reported by earlier XMM-Newton studies. As mentioned earlier, Sersic-159 is another cluster for which the LAB value for $N_H$ differed significantly from the Dickey and Lockman value: $N_H(\hbox{LAB}) = 1.15 \ti 10^{20}\hbox{\,cm}^{-2}$ compared to $N_H(\hbox{DL}) = 1.8 \ti 10^{20}\hbox{\,cm}^{-2}$.
Chandra data on the same cluster was considered in \cite{11070932} using the revised LAB value for the absorption column, again finding a soft excess.

\subsubsection{Summary}
In summary, soft excess emission has consistently been found with many different satellites across many different clusters. While calibration and background subtraction is difficult, the consistency between different instruments with very different sources of systematic error
strongly suggests that the effect is real and not an instrumental artefact. The precise energy range of the excess is unclear. Some XMM-Newton studies
suggest that the excess reaches up to $1\keV$, but this is also a satellite where background and calibration issues are more pronounced.

We will base our analyses in this paper on the ROSAT determinations of the soft excess and the numerical values given for the Coma cluster in the 38-cluster survey of
\cite{astroph0205473}. This analysis dates six years after the original discovery of the soft excess, allowing time for refinement of
analysis techniques and elimination of systematic errors, and it appears to be unchallenged.
ROSAT is also the best instrument for the study of the soft excess, and unlike the case for EUVE there is no controversy about
the precise mechanism of background subtraction.

%%%%%%%%%%%%%%%%%%%%%%%%%%%%%%%%%%%%%%%%%%%%%%%%%%%%%%%%%%%
%%%%%%%%%%%%%%%%%%%%%%%%%%%%%%%%%%%%%%%%%%%%%%%%%%%%%%%%%%%

\subsection{Astrophysical models of the soft excess}

Two main astrophysical models have been proposed to explain the soft excess. The first is the existence of a `warm' gas, which coexists with the hot gas of the ICM. In this scenario, the soft excess arises from thermal bremsstrahlung from this warm gas. The second scenario involves the non-thermal generation of the soft excess from inverse Compton scattering of relativistic electrons on the CMB.  Here we review both proposed explanations and note the observational constraints on them.

%%%%%%%%%%%%%%%%%%%%%%%%%%%%%%%%%%%%%%%%%%%%%%%%%%%%%%%%%%%

\subsubsection{The soft excess from a warm gas}
At the initial discovery of the soft excess it was noted that it may be explained by
an additional `warm' (but not hot) ICM component with temperature
$T \sim 0.5 $--$ 1 \ti 10^6\hbox{\,K}$ \cite{Lieu1996a, Lieu1996b, Bowyer96}.  In this scenario, the soft excess arises from thermal bremsstrahlung of the warm gas component, just as the dominant X-ray luminosity arises from thermal bremsstrahlung of the hot ICM gas.

This scenario is now generally regarded as problematic, particularly in the centre of clusters. It has two problems.
The first is that
a warm gas is only stable if it has comparable pressure to the hot gas of the ICM.
However, as $P = n k T$ and $T_\mathrm{warm} \approx 10^{-2} T_\mathrm{hot}$, this requires $n_\mathrm{warm} \approx 10^2 \, n_\mathrm{hot}$. Since the collisional cooling time goes as $n^{-2}$,
the warm gas is at the peak of its cooling curve and therefore cools rapidly, with a lifetime $t \approx 10^8\hbox{\,yrs}$ --- much shorter than the age of the cluster \cite{Fabian1996a, Fabian1996b}. The warm gas is then unstable against the rest of the cluster.

The second problem is that a
warm gas should also have thermal emission lines associated to it, in particular those associated with ${\rm O}_{\rm VI}$ and ${\rm O}_{\rm VII}$. However, searches for the far-ultraviolet ${\rm O}_{\rm VI}$  line predicted by this model have been null \cite{astroph9607126, astroph0102093}. At one point there were claimed detections
of  ${\rm O}_{\rm VII}$ lines \cite{astroph0210684, astroph0305424, astroph0309019}, but these claims were not supported by subsequent analysis, and the lines were found to be consistent with background and galactic foreground \cite{astroph0501007, astroph0602527, astroph0610461}.
While we note that there remains one marginal detection (approximately 3 sigma) of ${\rm Ne}_{\rm IX}$ and ${\rm O}_{\rm VIII}$ emission/absorption lines from Coma that could arise from a warm gas \cite{astroph0610424}, in general
further searches for emission lines have proved null \cite{astroph0608072, 07040475, 10103840, 11104422}, and in some cases this implies that any hypothetical warm gas must have a metallicity $Z < 10^{-3}$ \cite{11104422}.

While these two problems render the warm gas proposal  at the cluster centre very problematic,
it remains a possibility at the cluster outskirts where it has been suggested that the large soft excess halo around the outskirts of the
Coma cluster could originate from Warm-Hot Intergalactic Medium (WHIM) filaments. This proposal is discussed in e.g. \cite{astroph0409661, astroph0409707, 09033066}.

%%%%%%%%%%%%%%%%%%%%%%%%%%%%%%%%%%%%%%%%%%%%%%%%%%%%%%%%%%%

\subsubsection{The soft excess from  inverse Compton scattering}
In the second scenario, proposed in \cite{Hwang97, astroph9709232, astroph9712049, astroph9804310},
the soft excess arises from inverse Compton up-scattering of CMB photons with a relativistic electron population, hence abbreviated as IC-CMB. In the IC-CMB scenario, the average energy of a scattered photon is \cite{Longair}
\be
\langle E_\mathrm{scattered} \rangle = \frac{4}{3} \gamma^2 E_\mathrm{init} \, ,
\ee
for scattering off an electron with Lorentz factor $\gamma$. It follows that up-scattering CMB photons ($\langle E \rangle_\mathrm{CMB} \approx 3 k T_\mathrm{CMB} \approx 10^{-3}\hbox{\,eV}$) to the soft X-ray regime ($E \approx 200\hbox{\,eV}$) requires a population of non-thermal electrons with
\bea
\gamma \sim 500 \left( \frac{E_\mathrm{excess}}{200\hbox{\,eV}} \right)^{\half} \, . \nonumber
\eea
Such electrons have presumably been produced by supernovae, radio galaxies or by particle creation in intracluster shocks, and the IC-CMB explanation therefore ties the soft excess to the non-thermal cosmic ray content of clusters.

One attractive feature of this scenario is that significant populations of non-thermal relativistic electrons are known to be
present in some galaxy clusters. In Coma, for example, this is evidenced by the presence of a large radio halo which (as we will review in section \ref{sec:B}) indicates the presence of non-thermal relativistic electrons with $\gamma \approx 2000$. In the first versions of the IC-CMB scenario, the electrons at $\gamma \approx 500$ were assumed to connect
with the higher energy electrons at $\gamma \approx 2000$ by a simple spectral power law.

We now review in some detail how the IC-CMB explanation fixes the electron density at $\gamma \approx 500$ and how recent observations constrain the scenario.
For an IC-CMB origin of the soft excess,
 the number of relativistic electrons generating the excess can be determined by matching the radiation energy loss of the electrons to the observed soft X-ray excess. In detail, a relativistic electron loses energy (e.g. see \cite{astroph9901061}) as
\be
\frac{d \gamma}{dt} = - b(\gamma, t) \, ,
\ee
with $b= b_\mathrm{synchrotron} +b_\mathrm{IC-CMB}+b_\mathrm{Coulomb}+ b_\mathrm{bremsstrahlung}$, where for $\gamma \gg 1$,
\bea
b_\mathrm{synchrotron} & = & \frac{4}{3}\frac{\sigma_\mathrm{T}}{m_e c^2} \gamma^2 U_\mathrm{B} = 1.3 \ti 10^{-21} \gamma^2 \left( \frac{B}{1\,\mu\mathrm{G}} \right)^2 \hbox{s}^{-1}\, , \nn \\
b_\mathrm{IC-CMB} & = & \frac{4}{3}\frac{\sigma_\mathrm{T}}{m_e c^2} \gamma^2 U_\mathrm{CMB} = 1.37 \ti 10^{-20} \gamma^2 (1+z)^4 ~\hbox{s}^{-1}\, , \nn \\
b_\mathrm{Coulomb} & = & 1.2 \ti 10^{-15} \left( \frac{n_e}{10^{-3}\hbox{\,cm}^{-3}} \right) \left[ 1.1 + \frac{1}{75} \ln \left(
\gamma \cdot \frac{10^{-3}\hbox{\,cm}^{-3}}{n_e} \right) \right]~\hbox{s}^{-1}\, , \nn \\
b_\mathrm{bremsstrahlung} & = &1.5 \cdot 10^{-19} \left(\frac{n_e}{10^{-3}\hbox{\,cm}^{-3}}\right) \gamma \Big(0.36 + \ln(\gamma)  \Big)~~\hbox{s}^{-1}\, .
\eea
Here $\sigma_\mathrm{T}$ is the Thomson cross section, $m_e$ is the electron mass, $c$ is the speed of light and
 the factor of $(1+z)^4$ in $b_\mathrm{IC-CMB}$ takes into account the time-dependence of the CMB energy density.

The bremsstrahlung contribution arises from collisions of the relativistic electrons with the thermal ICM electrons and ions, here evaluated for  $y({\rm He})/y({\rm H})=0.1$. We note that $b_{\rm bremsstrahlung} < b_{\rm Coulomb}$ for $\gamma \lesssim 10^3$, and that  $b_\mathrm{bremsstrahlung} < b_\mathrm{IC-CMB}$ for $\gamma \gtrsim 100$ and $n_e \approx 10^{-3}\mathrm{\,cm}^{-3}$, so we may consistently neglect $b_\mathrm{bremsstrahlung}$.

 For sufficiently large magnetic fields the synchrotron losses dominate over the IC-CMB losses, and at $z=0$
 we find
 \be
 \frac{b_\mathrm{synchrotron}}{b_\mathrm{IC-CMB}} \eqsim \left(\frac{B}{3.2\,\mu{\rm G}}\right)^2 \, .
 \ee
 The Coulomb losses dominate over both IC-CMB and synchrotron losses at small enough $\gamma$,
 \be
 \frac{b_\mathrm{synchrotron}}{b_\mathrm{Coulomb}}  = \left(\frac{\gamma}{200}\right)^2 \left(\frac{B}{5\,\mu{\rm G}}\right)^2 \left( \frac{10^{-3}\,{\rm cm}^{-3}}{n_e}\right) \, ,
 \ee
 but become sub-leading for $\gamma \gtrsim 200$.

In the IC-CMB scenario, the soft excess arises from the energy loss of the relativistic electrons in the inverse Compton channel, for which the fractional energy loss is
\be
\frac{1}{\gamma} \frac{d \gamma}{d t}\Big|_{\rm IC-CMB~only} = - 6.9 \ti 10^{-18} \left( \frac{\gamma}{500} \right) ~\hbox{s}^{-1} \, .
\ee
Taking the Coma cluster as an example, we note that this fractional energy loss should result in a luminosity of approximately $\mc{L}_\mathrm{EUV} \approx 10^{43}\hbox{\,erg s}^{-1}$ from the central $500\kpc$ region.
The energy stored in relativistic electrons should be well-approximated by
\be
E_{\rm stored} \sim 1.5 \ti 10^{60} \left( \frac{\mc{L}_\mathrm{EUV}}{10^{43}\hbox{\,erg s}^{-1}} \right) \left( \frac{500}{\gamma} \right) ~\hbox{erg}.
\ee
Since each relativistic electron has energy $ 4 \ti 10^{-4} \left( \frac{\gamma}{500} \right)\hbox{\,erg}$, the total number of relativistic electrons responsible for the soft excess is given by
\be
\mc{N}_{\rm electron} \sim 3.8 \ti 10^{63} \left( \frac{\mc{L}_\mathrm{EUV}}{10^{43}\hbox{\,erg s}^{-1}} \right) \left( \frac{500}{\gamma} \right)^2 \, .
\ee
This population then corresponds to an average number density in the central region of $n_e^{\rm rel.} \approx 10^{-10}\hbox{\,cm}^{-3}$.
In this simple estimate we have considered a fixed $\gamma$, but the electron population is more likely distributed between $\gamma \approx 300$ and $\gamma \approx 600$ (corresponding roughly to the energies required to
generate IC-CMB photons between 75 and $300\hbox{\,eV}$).

%%%%%%%%%%%%%%%%%%%%%%%%%%%%%%%%%%%%%%%%%%%%%%%%%%%%%%%%%%%

\subsubsection{Observational constraints on the IC-CMB scenario}
To establish the  plausibility of the IC-CMB scenario,
it is crucial to understand the origin and evolution of the relativistic electrons necessary for the scenario (dedicated studies
are given in \cite{astroph9901061, astroph9906272, astroph9912557,
astroph0008518, astroph0011301}). Let us summarise the key behavioural features.
Since $\frac{d E}{dt} \sim - E^2$, high energy electrons are rapidly
degraded. If there is a single injection event with an initial spectrum $N(E) \sim N_0 E^{-p}$, radiative losses remove
 the high energy tail and generate an exponential cutoff above a maximal energy $E_{\rm max}(t)$, which is a decreasing function of time. However, for a continuous power-law injection the radiative losses steepen the spectrum, but the overall shape remains a power law.

The most severe constraints on the
IC-CMB scenario
arise from additional emission in the synchrotron and bremsstrahlung channels from the relativistic electron population.
Most early studies of the soft excess
\cite{Hwang97, astroph9709232, astroph9712049, astroph9907125, astroph9906272, astroph0008518, astroph0011301, astroph0101145, astroph0108369}
aimed to join the soft excess IC-CMB electrons with the synchrotron-emitting electrons responsible for the Coma radio halo through a single power law. A general result of these earlier papers was that this is indeed possible --- provided the cluster magnetic field is $B \lesssim 1\,\mu \hbox{G}$. For example, one of the first IC-CMB papers \cite{Hwang97} found that extrapolation of the radio population to the
soft excess regime could account for a fraction $\sim \left( \frac{B}{0.4\,\mu{\rm G}} \right)^{-2.34}$ of the soft excess in the Coma cluster.
The strong dependence on the magnetic field is easily understood from the fact that as the magnetic field grows, there is a growth in synchrotron emissivity for a fixed
number of electrons; meanwhile the $\gamma$-factor required for observable radio emission moves closer to that required for soft emission.

Consequently, if the magnetic field is comparatively large $(B \approx 5\,\mu{\rm G})$, then the soft excess electrons and the radio electrons
cannot belong to the same population.\footnote{For example, this is pertinent for the IC-CMB explanation of the soft excess in the Coma cluster, in which the magnetic field in the central region has recently been estimated to $B \approx 4.7\,\mu$G, and between 3-7 $\mu G$ at 99\% confidence level (as further discussed in section \ref{sec:B}).} The consequences of high magnetic fields for IC-CMB models of the soft excess have been studied in
\cite{astroph9912557, TsayBowyer}. In this case, it is essential that there is a sharp exponential cutoff on the extension of the
soft excess population to higher energies. For example, with $B \approx 5\,\mu{\rm G}$ soft excess emission is generated at $\gamma \sim 300 - 500$,
while observable radio emission starts from $\gamma \approx 1000$. Any power-law extension of the $\gamma \sim 300 - 500$ population, even with a spectral break,
overproduces synchrotron emission at $\gamma \approx 1000$.
It is then necessary that a sharp exponential cutoff is generated
in the interval between these energies.

As mentioned above, exponential cutoffs in the non-thermal electron spectrum may be produced simply by the time evolution of an isolated initial injection event with a power-law spectrum. However, for the IC-CMB scenario,  this requires an injection event occurring within a rather specific time interval in the cluster's past. If the injection event occurred  too long ago, the electrons have cooled
below $\gamma \approx 300$ and no soft excess can be obtained. If the injection event happened too recently, the electrons have not cooled sufficiently and would populate the $\gamma \gtrsim 1000$ region, thus overproducing synchrotron radio emission. In \cite{TsayBowyer}, it was concluded that this explanation of the soft excess in Coma requires an injection event between 1 and 1.4 billion years in the past.  These models furthermore require additional smaller injection events
to generate the population of electrons responsible for the radio halo.
We expect similar constraints on the clusters' injection histories to apply for other clusters exhibiting soft X-ray excesses.

In earlier work on IC-CMB explanations for the soft excess \cite{astroph9901061}, it had been argued that soft emission from $\gamma \approx 300$ electrons was expected to be a generic feature of galaxy clusters, as this value of the $\gamma$ factor maximises the characteristic loss time, $t \approx 3 \ti 10^9\hbox{\,years}$. As also noted in \cite{astroph9901061}, this ceases to be true once $B \approx 5\,\mu{\rm G}$. The loss time is now maximised at $\gamma \approx 100$ (and at $t \approx 10^9\hbox{\,years}$) and $\gamma \sim 300 - 500$ is no longer a special value. This reinforces the conclusions of the previous
paragraph: high magnetic fields require rather special initial conditions to generate the large electron population necessary for generating the soft excess without generating excessive radio emission.

We note that the aforesaid conclusions become even stronger for the case of cool-core clusters with a soft excess, for example, Virgo or A1795. Since cool core clusters have high central magnetic fields $B \gtrsim 10\,\mu{\rm G}$, the radiative cooling time is very short and $\gamma \approx 300$ electrons
have lifetimes of $\approx 3 \ti 10^8\,\hbox{years}$, much shorter than cluster timescales. It then requires a considerable coincidence to `catch' these electrons while they do not produce radio emission but do produce soft excess emission.

A further constraint on the IC-CMB explanation of the soft excess comes from gamma ray emission. Gamma ray emission arises from non-thermal bremsstrahlung from scattering of the IC-CMB electrons
off the  thermal proton population, and  is independent of the magnetic field.
 Since it is physically implausible that whichever initial event accelerated the electrons
did not simultaneously accelerate protons, there is in addition expected to be gamma ray emission from $\pi_0$ secondaries produced by collision of cosmic-ray protons with the thermal protons (note that at the top of the atmosphere cosmic-ray ions outnumber electrons by a factor $\approx 10^2$).

Fermi-LAT represents a significant
improvement in sensitivity over the EGRET gamma-ray telescope. Galaxy clusters represented one of the targets for Fermi, with in many
cases an expectation that Fermi would observe gamma rays. These expectations have not been fulfilled, with so far no positive detections of diffuse gamma ray emission from the intracluster medium \cite{10060748, 12010753, 12076749, 13085654, 13086278, AndoZandanel}.

The $\gamma$-ray spectrum produced in the IC-CMB scenario has been studied for the Coma cluster in \cite{astroph9906272} and \cite{astroph9912557}. In both cases,
the expected emission was found to be significantly greater than the projected Fermi (formerly known as GLAST) limits. For example, \cite{astroph9906272} found a predicted flux $N(\gamma)_{E > 100 \, {\rm MeV}} = 2 \ti 10^{-8}\,\hbox{cm}^{-2} \,\hbox{s}^{-1}$ (similar results are found in
\cite{astroph9912557}). This contrasts with bounds from the Fermi 5-year data of $N(\gamma)_{E > 100 \,{\rm MeV}} < 6 \ti 10^{-10}\,\hbox{cm}^{-2} \,\hbox{s}^{-1}$ \cite{AndoZandanel} (for a central point source template).

As these bounds are derived assuming a power-law photon index, and 
as the Fermi sensitivity to photons with $E_{\gamma} \approx 100\,\hbox{MeV}$ is less than originally anticipated, it may be possible to
try to tune the spectrum, with a very sharp cutoff at $E > 100\,\hbox{MeV}$, to be compatible with the non-observation of gamma rays.
While such tuned initial conditions may potentially be possible for any single cluster, it is difficult to see how the IC-CMB scenario can reconcile both the generic presence of the soft excess phenomenon and the generic absence of gamma rays from clusters at the level accessible to Fermi-LAT.\footnote{Note that while for clusters
along sightlines with large $N_H$ their intrinsic soft emission is unobservable, the same would not be true of their associated gamma rays. Therefore, when considering
the implication of the soft excess as a general feature of galaxy clusters, we can include in our sample additional clusters for which soft emission
is observationally inaccessible.}

A further difficulty for the IC-CMB explanation of the soft excess is the observation of soft excess emission at large radii from the cluster
centre \cite{astroph0211439, 09033067}.
In the case of Coma, soft excess halo emission has been detected up to radii of $\approx 5 \hbox{\,Mpc}$. This emission is well
beyond the radius at which the hot gas can be detected and at which the cluster meaningfully exists.
For an IC-CMB explanation of this halo, it is  unclear where this necessary relativistic electron population would come from.

A couple of other modified explanations or variations have been proposed. In \cite{astroph9808139} it was proposed that the soft excess could be generated by IC-starlight instead of IC-CMB. A deficiency of this model is that the energy density of starlight is far less than that of the CMB, and so it requires a proportionately greater energy in relativistic electrons, which would in fact constitute the dominant source of pressure in the cluster.
This model also requires a magnetic field $B \lesssim 1\,\mu{\rm G}$, inconsistent with Faraday rotation measurements. In \cite{Bowyer2004} it was proposed
to generate the relativistic electrons used in IC-CMB as secondaries from inelastic cosmic ray collisions, rather than as
primaries accelerated to very high energies by supernovae or active galactic nuclei. However, it was argued in \cite{astroph0701592} that this explanation can be excluded, as the secondary origin of the electrons allows the number of relativistic protons to be determined, and these then have an energy content far greater than the thermal content of the cluster.

Overall, while it would be premature to conclude that astrophysical explanations cannot work, the above difficulties and observations
 motivate alternative scenarios.

%%%%%%%%%%%%%%%%%%%%%%%%%%%%%%%%%%%%%%%%%%%%%%%%%%%%%%%%%%%
%%%%%%%%%%%%%%%%%%%%%%%%%%%%%%%%%%%%%%%%%%%%%%%%%%%%%%%%%%%
%%%%%%%%%%%%%%%%%%%%%%%%%%%%%%%%%%%%%%%%%%%%%%%%%%%%%%%%%%%

\section{A Cosmic Axion Background} \label{sec:CAB}

A generic feature of the four-dimensional effective theories arising from compactifications of string theory is the presence of massive scalar particles with feeble, Planck-mass suppressed interactions. Such particles --- normally called moduli --- parametrise the size and shape of the compactification geometry and set the values of coupling constants in the four-dimensional effective theory. While there is no absolute prediction for the moduli masses, in models where supersymmetry is relevant to the weak hierarchy problem moduli are expected to be at most a few orders of magnitude heavier than the weak scale \cite{CoughlanRoss, hepph9308292, hepph9308325}.

If moduli exist, general arguments imply they should be responsible for the reheating of the Standard Model degrees of freedom. Almost independently of the detailed model of inflation,  moduli become displaced
from their final metastable minimum during inflation
and begin to oscillate at the end of inflation.\footnote{The displacement is driven by the large inflationary energy density and its coupling to the moduli fields.
 A large displacement will arise whenever $V_{\rm inf} \gtrsim m_{\Phi, {\rm vac}}^2 M_\mathrm{P}^2$, where $m_{\Phi,{\rm vac}}$ is the vacuum mass of the modulus, and in practice moduli domination will come to occur
 even for very small initial displacements.}
  An oscillating scalar field redshifts like matter,
 \begin{equation}
 \rho_{\rm moduli} \sim a(t)^{-3} \, , \nonumber
 \end{equation}
where $a(t)$ denotes the Friedmann-Robertson-Walker scale factor. Owing to their feeble, Planck-mass suppressed interactions,
the moduli are long-lived. The oscillating moduli fields subsequently come to dominate over any initial radiation,
which redshifts as $\rho_{\rm radiation}~\!\sim~\!a(t)^{-4}$.
The universe then enters a modulus-dominated stage, which lasts until the moduli decay
into visible and hidden sector matter and radiation, thus inducing reheating.

 The characteristic decay rate of a modulus of mass $m_{\Phi}$ is
\be
\Gamma \sim  \frac{1}{8 \pi} \frac{m_{\Phi}^3}{M_{\rm P}^2}\, ,
\ee
where $M_\mathrm{P}$ denotes the reduced Planck mass, $M_{\rm P} = 2.48 \ti 10^{18}$ GeV. The energy density of the universe at the time of modulus decay, $\tau_{\rm decay}^{-1} \sim H \sim \Gamma$, is
\be
V_{\rm decay} \sim 3 H^2 M_\mathrm{P}^2 \sim \frac{m_{\Phi}^6}{M_\mathrm{P}^2}\, .
\ee
The visible sector decays of the modulus rapidly thermalise and initiate the Hot Big Bang at a temperature
\be
T_{\rm reheat} \sim \frac{m_{\Phi}^{3/2}}{M_\mathrm{P}^{\half}} \sim 1\hbox{\,GeV} \left( \frac{m_{\Phi}}{10^6\hbox{\,GeV}} \right)^{3/2}.
\ee
However, the gravitational origin of the moduli --- for example as extra-dimensional modes of the graviton --- implies that moduli can also decay to any hidden sector.
Furthermore, visible and hidden sector decay modes are approximately democratic, and in particular
the branching ratios
into hidden sector massless particles with extremely weak interactions (such as axions)
 need not be vanishingly small \cite{12083562,12083563,13047987, Angus:2013zfa} (also see \cite{12072771}).

Two-body decays of a modulus field into axions are induced by the Lagrangian coupling $\frac{\Phi}{M_P} \partial_{\mu}a\partial^{\mu}a $,
resulting in axions with an initial energy $E_a = m_{\Phi}/2$. Since they are weakly interacting, the axions do not thermalise and the vast majority of axions propagate freely to the present day, where they form a homogeneous and isotropic Cosmic Axion Background. Furthermore, being relativistic, they contribute to the dark radiation energy density of the universe.

The characteristic axion energy today is set by the initial axion energy, redshifted to the present. Since the current CMB temperature is found simply by
redshifting the primordial thermal plasma (up to a small $\left( \frac{g_{*,{\rm now}}}{g_{*,{\rm init}}} \right)^{-1/3}$ boost as species decouple), we have
$$
\frac{E_{a,{\rm now}}}{T_{\gamma,{\rm now}}} \simeq \frac{E_{a,{\rm init}}}{T_{\gamma,{\rm init}}} \sim \left(\frac{M_\mathrm{P}}{m_{\Phi}} \right)^{1/2} \, .
$$
For moduli masses $m \approx 10^6\hbox{\,GeV}$, this gives $E_a \sim 10^6\,T_{\rm CMB} \approx 200\hbox{\,eV}$.

\begin{figure}[]
\begin{center}
\includegraphics[width=.65 \textwidth]{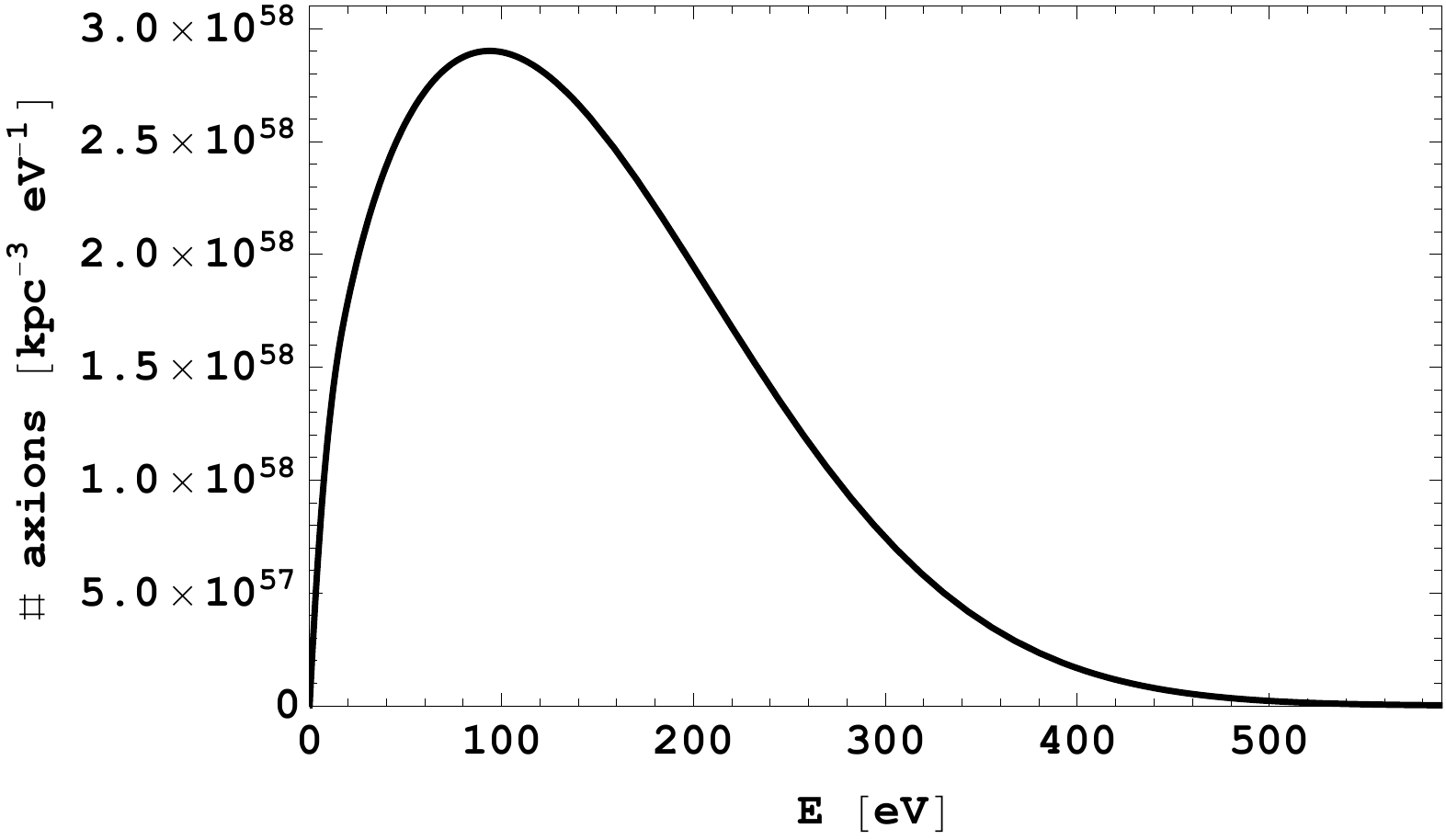}
\end{center}
\caption{A typical axion number density per (kpc)$^3$ for a CAB with $\langle E_{CAB}\rangle = 150$ eV, which contributes to  dark radiation with $\Delta N_{\rm eff} = 0.5$. The precise location of the energy peak depends on the value of $m_{\Phi}$.
 }
\label{fig:CAB}
\end{figure}

To find the exact spectral shape of the CAB, we must account for the fact that moduli do not decay instantaneously, and meanwhile the expansion rate of the
universe changes as it transitions from matter (modulus) domination prior to reheating into radiation domination after all moduli have decayed.
Moduli that decay early give rise to present-day lower-energy axions as they have more time to redshift, whereas more energetic axions arise from late-decaying
moduli. The spectral shape was computed numerically in \cite{13041804} and may be described as `quasi-thermal', with an exponential fall-off at high energies (c.f.~figure \ref{fig:CAB}).
The overall magnitude is normalised to the axionic contribution to $\Delta N_{\rm eff}$, and the peak location is determined by the mass of the modulus
and its lifetime.

While a CAB can be indirectly probed through studies of dark radiation,
it can be directly observed only through its couplings to visible-sector matter and gauge bosons, as mediated, for example, by the operator
$$
\frac{a}{M} {\bf E} \cdot {\bf B} \, .
$$
In the presence of a magnetic field this induces `oscillations' of axions into photons in a process analogous to neutrino oscillations \cite{Sikivie:1983, Sikivie:1985}. The observational consequences of this conversion of the CAB have been considered in
\cite{CAB, 13066518, 13104464}. Axions may also play a role by scattering off ambient particles in the thermal plasma, which was considered in \cite{13041804}.

Giving the value of $M$, the total axionic energy density %$\Delta N^{(a)}_{\rm eff}$ 
and the central CAB energy specifies an entirely predictive model. In this model,
the spectrum and number of photons arising from axion-photon conversion in any astrophysical magnetic field can be computed.

%%%%%%%%%%%%%%%%%%%%%%%%%%%%%%%%%%%%%%%%%%%%%%%%%%%%%%%%%%%
%%%%%%%%%%%%%%%%%%%%%%%%%%%%%%%%%%%%%%%%%%%%%%%%%%%%%%%%%%%

\subsection{What would a CAB tell us?}

We want to ask what could be learnt from the existence of a Cosmic Axion Background at a given energy $E_a$. We are going to assume that
this is generated by the primordial non-renormalisable decays of a field $\Phi$ of mass $m_{\Phi}$ and coupling constant $\Lambda$, decaying with
\be
\Gamma = \frac{1}{8 \pi} \frac{m_{\Phi}^3}{\Lambda^2} \, .
\ee
What would observations of a CAB tell us about $m_{\Phi}$ and $\Lambda$? The two points we use are the ratio $\frac{E_a}{T_{\rm CMB}}$ (assumed to be measured)
and the requirement that the
reheat temperature must be greater than the BBN temperatures.

We assume the instantaneous decay approximation, under which all $\Phi$ particles decay at a time
\be
\tau = \Gamma^{-1} = 8 \pi \frac{\Lambda^2}{m_{\Phi}^3} \, .
\ee
While continual decays give a more refined analysis,
the instantaneous decay approximation is here sufficient to capture the key physics.

As the universe is matter dominated until the point of decay, the Hubble constant at time of decay is $H_{\rm decay} = \frac{2}{3 \tau}$, with
\be
H(\tau) \equiv H_{\rm decay} = \frac{1}{12 \pi} \frac{m_{\Phi}^3}{\Lambda^2} \, .
\ee
We assume $\Phi$ decays to the visible sector with branching ratio $(1-B_a)$ and to the axion with branching ratio $B_a$. The initial Standard Model
energy density is
\be
\rho_{\rm SM} = (1- B_a) \ti 3 H_{\rm decay}^2 M_\mathrm{P}^2 \, .
\ee
We assume instant thermalisation so that
\be
\frac{\pi^2}{30} g_{*}(T_{\rm rh}) T_{\rm rh}^4 = (1-B_a) \ti 3 H_{\rm decay}^2 M_\mathrm{P}^2 \, .
\ee
This gives
\be
T_{rh} = \left( \frac{5 (1-B_a)}{8 \pi^4 g_{*}(T_{\rm rh}) } \right)^{1/4} \frac{m_{\Phi}^{3/2} M_\mathrm{P}^{1/2}}{\Lambda} \, .
\ee

The initial axion energy is $E_{a,{\rm rh}} = m_{\Phi}/2$. As the universe expands, the axion energies redshift directly as $R^{-1}$
whereas the photon energies redshift as $g_{*}^{-1/3} R^{-1}$. We then have
\be
\left( \frac{T_{\gamma}}{E_a} \right)_{now} = \left( \frac{1}{10.75} \right)^{1/3} \left( \frac{11}{4} \right)^{1/3}
g_{*}(T_{\rm rh})^{1/3} \left( \frac{T_{\gamma}}{E_a} \right)_{\rm reheat}.
\ee
This takes into account that there are two distinct boosts to the photon temperature as species become non-relativistic: one from the time of reheating to the time of neutrino decoupling, and the second from the time of neutrino decoupling to the present.
We evaluate this to find
\be
\left( \frac{E_a}{T_{\gamma}} \right)_{\rm now} = 2.78 (1 - B_a)^{-1/4} g_{*}^{-1/12}(T_{\rm rh}) \frac{\Lambda}{m_{\Phi}^{1/2} M_\mathrm{P}^{1/2}} \, .
\ee

If we suppose a CAB is measured, then we fix $\frac{E_a}{T_{\gamma}} = \lambda$ as a measured parameter. What does this tell us? Note that we can write
\be
T_{\rm reheat} = 0.36 \left( \frac{10.75}{g_{*}(T_{\rm rh})} \right)^{1/3} \frac{m_{\Phi}}{\lambda} \, .
\ee
We now approximate $g_{*} \approx 10.75, B_a \approx 0$, and impose $T_{\rm rh} > 3\hbox{\,MeV}$ for consistency with BBN. This gives
\be
m_{\Phi} > 10 \lambda\hbox{\,MeV},
\ee
or equivalently
\be
\Lambda \gtrsim 7 \ti 10^{16} \left( \frac{\lambda}{10^6} \right)^{3/2}\hbox{GeV} \, .
\ee

Thus the assumed observation of a CAB with energies in the few hundred eV range would imply the existence of an extremely weakly coupled particle whose
interactions are suppressed by a scale $\Lambda \gtrsim 7 \ti 10^{16}\hbox{\,GeV}$. If we also impose $M_\mathrm{P} \geq \Lambda$, then we can bracket the mass
and coupling of $\Phi$:
\bea
\left( \frac{10^6}{\lambda} \right)^2 1.2 \ti 10^7 \hbox{\,GeV} \gtrsim m_{\Phi} \gtrsim \left( \frac{\lambda}{10^6} \right) 10^4 \hbox{\,GeV} \, , \\
7 \ti 10^{16} \hbox{\,GeV} \lesssim \Lambda \lesssim M_\mathrm{P} \, . \label{eq:scale}
\eea
An observation of a CAB can then be used to also show the existence of heavy massive particles which interact only through couplings suppressed by a scale close to
the four dimensional Planck scale --- precisely the properties string compactifications predict for moduli.

%%%%%%%%%%%%%%%%%%%%%%%%%%%%%%%%%%%%%%%%%%%%%%%%%%%%%%%%%%%
%%%%%%%%%%%%%%%%%%%%%%%%%%%%%%%%%%%%%%%%%%%%%%%%%%%%%%%%%%%
%%%%%%%%%%%%%%%%%%%%%%%%%%%%%%%%%%%%%%%%%%%%%%%%%%%%%%%%%%%

\section{Cluster magnetic fields} \label{sec:B}

In this section we will discuss magnetic fields in galaxy clusters. In section \ref{sec:obs} we briefly review the observational methods used to infer the existence of cluster magnetic fields, and in section \ref{sec:ComaB} we review the magnetic field model of \cite{10020594}.

%%%%%%%%%%%%%%%%%%%%%%%%%%%%%%%%%%%%%%%%%%%%%%%%%%%%%%%%%%%
%%%%%%%%%%%%%%%%%%%%%%%%%%%%%%%%%%%%%%%%%%%%%%%%%%%%%%%%%%%

\subsection{Magnetic fields in galaxy clusters} \label{sec:obs}

It is by now well-established that galaxy clusters  support magnetic fields with typical field strengths of ${\cal O}(B) \sim 1$--$10\,\mu$G.
Such fields have presumably arisen from the exponential amplification of much smaller `seed' magnetic fields through some dynamo mechanism, but little is known for certain about either the origin of the seed fields or the mechanism of amplification.
As reviewed in \cite{Widrow:2002}, if not present primordially, seed magnetic fields may have been generated from astrophysical processes such as those of active galactic nuclei (AGN).

While there is similarly no consensus regarding the  mechanism for amplification of cluster magnetic fields, it is clear that the $\alpha\Omega$-dynamo, which may be responsible for the amplification of galactic magnetic fields, is not active in galaxy clusters. The $\alpha\Omega$-dynamo amplifies magnetic fields though the conversion of mechanical energy to magnetic energy in the turbulent and differentially rotating motion of the galaxy disc, but
becomes ineffective in systems such as clusters with little or no angular rotation.
`Small-scale' dynamo mechanisms in which ICM turbulence leads to repeated shearing of the magnetic field and to a subsequent explosive amplification of the field strength have been suggested, but cluster magnetic field amplification remains an active research topic.

However, observational studies of cluster magnetic fields are making steady progress, and here we briefly review the methods used to infer their existence and the resulting estimates for the magnetic field in the Coma cluster.

The first evidence for the existence of cluster magnetic fields was obtained from observations of the radio halo of the  Coma cluster.
This large region of diffuse radio emission extends to radii $>1\,\mathrm{Mpc}$ from the  centre  of the cluster. The radiation from the halo cannot plausibly
be associated with the integrated luminosity of the constituent galaxies --- the only viable explanation is that it is synchrotron radiation from a
diffuse population of relativistic electrons in the cluster magnetic field \cite{Willson:1970}.

 For  the Coma radio halo,  large-scale diffuse emission across the cluster
has been measured at frequencies between $30\,$MHz and $4.5\,$GHz (for example see \cite{Henning89, Kim90, Giovannini1993}).
An electron with boost factor $\gamma$ in a magnetic field $B$ emits synchrotron radiation at a frequency \cite{Longair}
\be
\nu \sim 4  \gamma^2 \left( \frac{B}{ 1 \mu \hbox{G} } \right) \hbox{\,Hz} \, .
\ee
The radio halo then requires a population of electrons between $\gamma_\mathrm{min} \simeq \left( \frac{5\,\mu\mathrm{G}}{B} \right)^{\half} 1.2 \ti 10^3$ and \hbox{$\gamma_\mathrm{max} \simeq \left( \frac{5\,\mu\mathrm{G}}{B} \right)^{\half} 1.5 \ti 10^4$}.
An electron number density $N(E) \sim N_0 E^{-p}$ generates a synchrotron emissivity \cite{Longair}
\be
J(\nu) \sim N_0 B^{(p+1)/2} \nu^{-\alpha},
\ee
with $\alpha = (p-1)/2$.
The Coma radio halo has an apparent break around $0.6\,$GHz: below this frequency the synchrotron index is $\alpha \approx 1$, while above this frequency
$\alpha \approx 2$. These values correspond to electron number indices $p \approx 3$ or $p \approx 5$. If a single index is fitted across the radio spectrum, then we obtain an electron spectrum $p \approx 3.5$.

The total level of synchrotron emission correlates with the strength of the magnetic field, and the degree of polarisation serves as an indicator of field uniformity and structure.
However, the actual magnitude of the magnetic field cannot be determined absolutely from synchrotron emissivity, due to the degeneracy with the size of the electron population.
An estimate of the magnetic field strength can be made using `equipartition' arguments, in which the total energy content of the synchrotron-emitting relativistic particles and magnetic field is minimised.  Since such arguments are always based on assumptions that may not be easily verifiable, equipartition arguments for magnetic field strengths can only give a rough indication of the average magnetic field across the radio halo.  With this method, the radio emission may be attributed to the large-scale distribution of non-thermal relativistic electrons of GeV energies (i.e.~with $\gamma \approx 2000$) subject to $\approx 1\,\mu\mathrm{G}$ magnetic fields.  In particular, in reference \cite{Thierbach:2003} the Coma magnetic field averaged over
the central $1\hbox{\,Mpc}^3$ was estimated to be $B \sim 0.7-1.9\,\mu$G based on equipartition arguments.

The relativistic electrons responsible for the Coma radio halo will also inverse-Compton (IC) scatter off CMB photons to produce hard X-ray photons
with frequencies around
\be
\nu_\mathrm{IC} = \frac{4}{3} \gamma^2 \nu_0 \sim \left( \frac{\gamma}{2000} \right)^2 4\hbox{\,keV} \, .
\ee
More energetic electrons would similarly scatter CMB photons into gamma-rays. Thus in principle an observation of a hard X-ray inverse-Compton signal
could be used to break the degeneracy between electron density and magnetic field size, allowing for a direct determination of the cluster magnetic field. In reference \cite{Femiano:2004} this argument was used to estimate the Coma magnetic field as $B \approx 0.2\,\mu$G. However, more recent analysis of hard X-ray observations by numerous experiments shows no evidence for a non-thermal hard X-ray component in Coma, in which case the IC method only leads to lower limits on the magnetic field strength: $ B > 0.2\,\mu\hbox{G}$ for the Coma radio halo,
and $B> 1\,\mu$G for the Coma radio relic \cite{Chen:2008}.

A fundamentally different method for estimating cluster magnetic fields comes from Faraday rotation.  The magnetised ICM plasma induces different phase velocities for right-handed and left-handed photons and thus becomes birefringent. For linearly polarised light produced from e.g.~synchrotron emission from localised radio sources, this leads to an effective rotation of the plane of polarisation of the wave as a function of wavelength, called Faraday rotation.  This effect is conveniently estimated in terms of a `rotation measure', RM, defined by
\be
\Psi_\mathrm{obs}(\lambda) = \Psi_{0} + \lambda^2\,\mathrm{RM} \, .
\ee
The rotation measure is given by the
 the line-of-sight integral of the parallel component of the magnetic field multiplied by the electron density,
\be
\mathrm{RM} = \frac{e^3}{2\pi m_e^2} \int_{\rm l.o.s.} n_e(l) B_{\parallel}(l) dl \, ,
\ee
where, by convention, a magnetic field pointing towards the observer gives rise to a positive RM.
Since the electron density distribution may be inferred from X-ray observations of the thermal ICM, studies of rotation measures provide a sensitive probe of the magnitude of the cluster magnetic field.

Thus, in principle, by measuring $\Psi_\mathrm{obs}$ at several frequencies for a given radio source, the value of RM may be inferred. The rotation measures typically exhibit a patchy structure across a radio source, and by studying the statistics of the RM distributions the scale over which the ICM magnetic field becomes tangled can be estimated.  Finally, by considering a number of radio sources emitting linearly polarised light in and behind a  galaxy cluster, the magnitude of the magnetic field as well as its radial dependence may be estimated. For a recent review, see \cite{Feretti:1205.1919}.\footnote{For a contrary view arguing that the rotation measure is attributed to the source radio galaxy and not the
ICM as a whole, see \cite{RudnickBlundell2003}.}

A  number of studies of Faraday rotation measures in the Coma cluster have been performed.  In \cite{Kim:1990}, 18 radio sources were analysed and a significant enhancement of the RM towards the centre of the cluster was found. To estimate the magnetic field strength, a simple model of the magnetic field reversal was used in which the magnetic field and electron density were assumed to be constant in magnitude throughout the ICM, but the magnetic field orientation was assumed to perform a random walk with a fixed step size $\Lambda_B$, corresponding to the magnetic field autocorrelation length. Such a model always results in $\langle \mathrm{RM} \rangle =0$, but with a variance proportional to the square of the magnetic field strength,
 \be
 \sigma_\mathrm{RM}^2 = 812 \Lambda_B  \int_{\rm l.o.s.} \left(n_e(l) B_{\parallel}(l)\right)^2 dl  \, ,
 \ee
 in units of radians$^2$/m$^4$.  Using this model, the magnetic field strength was estimated as $B \approx 2\,\mu$G, with a tangling scale of $13-40\,$kpc.

In \cite{Feretti:1995}, RMs from the extended radio galaxy NGC 4869 in the Coma cluster were analysed, resulting in a field strength estimate of $B\lesssim 8.5\,\mu$G. Furthermore, the scale over which the magnetic field changes orientation was inferred to be $<1\,$kpc. Since the mean of the rotation measure from the radio source was shown not to vanish, the simple random walk model was amended with a constant component of strength $\approx 0.2\,\mu$G uniform over $\approx 200\,$kpc.

More recently, better but computationally more expensive
software tools have been designed to constrain the cluster magnetic field by simulating mock RM images from a magnetic field with a given power spectrum and by comparing the results to RMs obtained from radio observations \cite{Murgia:2004, LaingMurgia:2008}. These allow a treatment of multi-scale magnetic fields. As we will review in section \ref{sec:ComaB},
there are still some parameter degeneracies that complicate the final interpretation of this type of analysis, but the central magnetic field in Coma can
be constrained to $B\sim 3-7\,\mu$G.   %, and here we simply note that such a method will .
 An alternative semi-analytic method utilising Bayesian inference has also been developed \cite{EnsslinVogt:2004}.

In sum, observational evidence for cluster magnetic fields has been obtained by several independent methods, with observations of RMs from Faraday rotation giving the most direct estimate. These methods involve different theoretical assumptions and measure slightly different quantities (for example, average magnetic field in the case of synchrotron luminosity
versus line-of-sight magnetic field for Faraday rotation). It is therefore not surprising that the resulting estimated field strength can differ by a factor of a few.

%%%%%%%%%%%%%%%%%%%%%%%%%%%%%%%%%%%%%%%%%%%%%%%%%%%%%%%%%%%
%%%%%%%%%%%%%%%%%%%%%%%%%%%%%%%%%%%%%%%%%%%%%%%%%%%%%%%%%%%

\subsection{Coma magnetic field model} \label{sec:ComaB}
In this section we review the stochastic Coma magnetic field model of \cite{10020594}, which is based on the approach first proposed in \cite{Murgia:2004}. In this model, the magnetic field in the Coma cluster is simulated as a multi-scale, tangled
 field  with a field strength that scales with the electron density of the cluster. For certain model parameters,  mock RMs derived from this magnetic field model were shown to be in good agreement with the RMs of radio sources in the central Coma cluster region observed with the Very Large Array (VLA)
%, Bonafede at al. (2010)
\cite{10020594}. We note that while this model is a more sophisticated refinement of earlier models of the cluster magnetic field, it is still a model and cannot be expected to fully capture all features of a turbulent cluster magnetic field. Rather, this model presents a tractable approximation of the cluster magnetic field that has been shown to successfully reproduce some quantitative features of the true cluster field.

In the  stochastic  model of  \cite{10020594},  the  cluster magnetic field
is constructed with a specified power spectrum through the generation of a random vector potential with a power spectrum
\beq
\langle |\tilde A_k|^2 \rangle \sim \vert k \vert^{-n} \, .
\eeq
Such a
 vector potential  may be constructed in  momentum space  by randomly drawing  the magnitude of each component from  a Rayleigh distribution as, %of
\beq
	p(\tilde{A}_k)=\frac{\tilde{A}_k}{\vert k \vert^{-n}} \exp{ \left( -\frac{\tilde{A}_k^2}{2\vert k \vert^{-n}} \right) } \, ,
\eeq
where we have suppressed the vector index on $\tilde A_k$.
The complex phase of each Fourier component of the vector potential is taken to be  uniformly distributed between $0$ and $2\pi$.
The momentum space magnetic field is then calculated as
\beq
	\tilde{B}(k)=ik\times \tilde{A}(k) \, %,
\eeq
and has a power-law power spectrum
\beq
	P_{\tilde B} = \frac{1}{(2 \pi)^3} \langle |\tilde B(k)|^2 \rangle \sim \vert k \vert^{-n+2} \, . \label{eq:Bspectrum}
\eeq
A tractable numerical simulation of this field requires a truncation of the power spectrum in the IR and UV, i.e.~a restriction of the  momenta to some  range
\beq
\label{eq:kconstraint}
	k_{\mini} \leq k \leq k_{\maxi} \, ,
\eeq
where in real space this corresponds to fields with structure only on scales larger than  $\Lambda_{\mini} = 2\pi/k_{\maxi}$ and smaller than $\Lambda_{\maxi}=2\pi/k_{\mini}$. The `tangled' position-space magnetic field --- which will automatically be divergence-free with normally distributed components, $B_{\rm gen.}^i(x)$ --- may be obtained by Fourier transformation of $\tilde B(k)$. The real-space variance of the generated magnetic field is then as usual given by $\sigma^2_i = \frac{1}{2 \pi^2}\int_{k_{\mini}}^{k_{\maxi}} dk k^2 P_{\tilde B^i}$.

While the generated magnetic field $\vec{B}(x)$ exhibits structure over a range of scales, it does not reproduce one of the key properties of cluster magnetic fields --- the attenuation of the field strength with radius. Following \cite{10020594}, we note that such an attenuation can be modelled	
by enforcing that the magnetic field scales
with  the gas density of the hot intra-cluster medium. The density of the ICM in the central region of
 Coma is well-described by the $\beta$-model,
\beq
	n_e(r)=n_0\left(1+\frac{r^2}{r_c^2}\right)^{-\frac{3}{2}\beta}, \label{eq:betamodel}
\eeq
where the central electron density $n_0$, the core radius $r_c$, and $\beta$ have been %experimentally
determined %by Briel et al.
 to be \hbox{$3.44\ti 10^{-3}\hbox{cm}^{-3}$}, $291\kpc$, and $0.75$ respectively \cite{Briel:1992}.

A more realistic model of the magnetic field may then be obtained by
%We then
modulating $B^i_{\rm gen.}(x)$ by multiplication by some function of the intracluster electron density, $f(n_e)= {\cal C} B_0 \left(\frac{n_e(r)}{n_0}\right)^\eta$. The constant ${\cal C}$ is chosen so as to normalise the average magnetic field across the core of the cluster to some parameter value $B_0$.  The values $\eta$ and $B_0$ are then two additional parameters of the model. By comparing simulated  RMs from the above model  with rotation measures inferred from VLA observations, the authors of %reference
\cite{10020594} found the best fit values  $\eta=0.5$ and $B_0=4.7\mug$, but values in the ranges $\eta = 0.4$, with $B_0=3.9\mug$ to
$\eta=0.7$, with $B_0=5.4\mug$ gave fits within $1\sigma$ of observations.

We note that the total cluster magnetic field,
\be
\vec{B}^{\rm tot.} := {\cal C} \cdot B_0 \cdot \left(\frac{n_e(r)}{n_0}\right)^\eta \vec{B}_{\rm gen.}(r, \theta, \phi) \, , \label{eq:Btot}
\ee
is no longer divergence-free but receives local contributions from a fictitious magnetic monopole density proportional to $\vec{\nabla} n_e \cdot \vec{B}_{\rm gen.}$. This contribution, however, is always proportional to $B_0 \frac{L}{r_c}$, where $L$ is the coherence length of the magnetic field, and is negligible for  $L\ll r_c$.

It is important to note that while the parameters $n$, $\Lambda_{\mini}$, $\Lambda_{\maxi}$, $\eta$ and $B_0$ %are free parameters of the magnetic field model which
can be constrained by fitting to rotation measures  \cite{10020594},  there is an effective degeneracy between $n$ and $\Lambda_{\maxi}$. Larger $n$ can be compensated for by lowering $\Lambda_{\maxi}$ and vice versa, giving an equally good fit when comparing to RMs from  Faraday rotation.
The value $n=17/3$ corresponds to a Kolmogorov-like turbulent power-law slope for the one-dimensional power spectrum of the magnetic field (here defined as \hbox{${\cal P}(k) \sim 2\pi k^2|\tilde B_k|^2 \propto k^{-n+4}$}) and was the headline value adopted in \cite{10020594}. This power-law slope corresponds to a best fit value of $\Lambda_{\maxi} = 34\kpc$, with $\Lambda_{\mini}$ found to be $2\kpc$.
However, as discussed in \cite{10020594}, the Faraday rotation measurements are degenerate along a curve in $(n, \Lambda_{max})$ space, with a flatter spectrum as
$\Lambda_{max}$ is increased.
Equally good fits to Faraday rotation measures are provided by a flat one-dimensional power spectrum, i.e. $n=4$, with $\Lambda_{\max}$ increased to $100\kpc$.
These spectra have more power on small scales compared to the Kolmogorov spectrum.

In section \ref{sec:analysis} we will use these models of the Coma magnetic field to study conversion of axions in the cluster magnetic field.

%%%%%%%%%%%%%%%%%%%%%%%%%%%%%%%%%%%%%%%%%%%%%%%%%%%%%%%%%%%
%%%%%%%%%%%%%%%%%%%%%%%%%%%%%%%%%%%%%%%%%%%%%%%%%%%%%%%%%%%
%%%%%%%%%%%%%%%%%%%%%%%%%%%%%%%%%%%%%%%%%%%%%%%%%%%%%%%%%%%

\section{The soft excess in Coma from the CAB} \label{sec:analysis}

This section brings together the topics reviewed in the above sections.
We now describe the
implementation and results of a numerical simulation of axion-photon conversion of the CAB in magnetic fields generated stochastically using the
magnetic field models described in section 4. %and a comparison to the experimental data.
In the results section \ref{sec:results} we discuss both particular and generic features of the generated photon flux, and we
compare our results for axion-photon conversion of the CAB to the observed soft excess luminosities in Coma \cite{astroph0205473}.

%%%%%%%%%%%%%%%%%%%%%%%%%%%%%%%%%%%%%%%%%%%%%%%%%%%%%%%%%%%
%%%%%%%%%%%%%%%%%%%%%%%%%%%%%%%%%%%%%%%%%%%%%%%%%%%%%%%%%%%

\subsection{Numerical simulation}
The simulation of axion-photon conversion in the Coma cluster can be divided into three steps: first, a stochastic magnetic field of the type consistent with observations of
Faraday rotation measures is generated on a large lattice; second, an initial axion state of a particular energy is quantum mechanically propagated  through this lattice; and finally, by normalising the initial axion distribution to the CAB spectrum derived in \cite{13041804}, the resulting photon luminosities and spectrum are obtained.

%%%%%%%%%%%%%%%%%%%%%%%%%%%%%%%%%%%%%%%%%%%%%%%%%%%%%%%%%%%

\subsubsection{Magnetic field generation} \label{sec:Bsimul}
Following the detailed prescription reviewed in section \ref{sec:ComaB},
we have generated a numerical model of the Coma magnetic field, on a $2000^3$ grid with an $s = 0.5\kpc$ unit cell size, using C++. %. The unit size of the magnetic field grid must needs be chosen so as to satisfy
This way the Nyquist criterion that the sampling rate of a dataset must be greater (ideally much greater) than twice the frequency of the dataset is satisfied for fields with structure only on scales larger than $\Lambda_{\mini} > 2s = 1\kpc$. We note that such a small unit cell size places a limit on the size of the field we can generate, making it impractical to go beyond $\approx1\Mpc^3$.

As outlined in section \ref{sec:ComaB}, the values of the Fourier coefficients of the vector potential %Field values
are generated randomly for all modes in the range of equation (\ref{eq:kconstraint}).
After computing the momentum space magnetic field, the real space representation is obtained by performing a discrete Fourier transform %of the magnetic field is performed
using {\sc FFTW 3.3.3} routines  \cite{FFTW}.

The real-space magnetic field is modulated as in equation (\ref{eq:Btot}) so as to exhibit attenuation  over cluster scales.
 The normalisation constant ${\mathcal C}$ is chosen so that the average magnitude of the magnetic field within the core radius, $r_c$, of the cluster is equal to the parameter $B_0$. In detail, this gives
\be
{\cal C} = \frac{N_{r<r_c} }{\sum_{r<r_c} B_{\rm gen.}(\frac{n_e}{n_0})^{\eta}} \, ,
\ee
where $N_{r<r_c}$ denotes the number of lattice points at radii less than the cluster radius.

As discussed in section \ref{sec:ComaB}, the observation of rotation measures from Faraday rotation does not completely determine the parameters of the stochastic model, but rather restricts their values to certain degeneracy classes. In this paper we consider three sets of magnetic field parameters, which are listed in table \ref{tab:params} on page
\pageref{tab:params}.

%%%%%%%%%%%%%%%%%%%%%%%%%%%%%%%%%%%%%%%%%%%%%%%%%%%%%%%%%%%

\subsubsection{Axion-photon propagation I: Homogeneous solution}
\label{sec:prop}

In this section we outline the theory  of axion-photon conversions in an external magnetic field. In the presence of an external magnetic field, axions and photons mix via the term
\beq
	\mathcal{L}\supset \frac{1}{8 M}a F_{\mu\nu}\tilde{F}^{\mu\nu} \equiv \frac{1}{M}a\vec{E}\cdot\vec{B} \, \equiv g_{a\gamma\gamma} a \vec{E} \cdot \vec{B} \, .
\eeq
From the wave equation for particles propagating in the $z$-direction, the corresponding linearised equation of motion for the axion-photon system is  \cite{Raffelt}
\beq
\label{eq:EqofMotion}
\left(\omega + \left(\begin{array}{c c c}
			\Delta_{\gamma} & \Delta_{\rm F} & \Delta_{\gamma a x} \\
			\Delta_{\rm F} & \Delta_{\gamma} & \Delta_{\gamma a y} \\
			\Delta_{\gamma a x} & \Delta_{\gamma a y} & \Delta_{a}
		   \end{array}\right) - i\partial_z\right)\left(\begin{array}{c}
								\Ket{\gamma_x} \\
								\Ket{\gamma_y} \\
								\axion
							      \end{array}\right)= 0 \, .
\eeq
Here $\omega$ denotes the energy of the  photon and axion modes and $\Delta_{\rm F}$ denotes  the  Faraday rotation of photon polarisation states due to the cluster magnetic field.
Since this mixing is between photons only, in the limit of small axion-photon mixing this effect is negligible and we will
henceforth set it to zero. The  refractive index for photons in the plasma is given by  $\Delta_{\gamma}  = -\omega_{\rm pl}^2/2\omega$, where $\omega_{\rm pl}  = \sqrt{\frac{4\pi\alpha n_e}{m_e}}$  denotes the plasma frequency of the ICM. The axion-photon mixing is induced by the matrix element $\Delta_{\gamma a i}  = B_i/2M$. The mass of the axion determines the final diagonal matrix element: $\Delta_a  = -m_a^2/\omega$.
Formally, we may write the general solution to equation (\ref{eq:EqofMotion}) as
\beq
\label{eq:homsoln}
\left(\begin{array}{c}
	\Ket{\gamma_x} \\
								\Ket{\gamma_y} \\
	\axion \\
\end{array}\right)(L) =  {\cal T}_z\left[ \exp \left(-i \omega L \mathbb{I} \ -i \int_0^L \mathcal{M}(z) \dd z\right) \right] \left(\begin{array}{c}
				\Ket{\gamma_x} \\
								\Ket{\gamma_y} \\
												\axion
		      \end{array}\right)(0) \, ,
\eeq
with
\beq
\mathcal{M} (z) = \left(\begin{array}{c c c}
							\Delta_{\gamma}(z) & 0 & \Delta_{\gamma a x}(z) \\
							0 & \Delta_{\gamma}(z) & \Delta_{\gamma a y}(z) \\
							\Delta_{\gamma a x}(z) & \Delta_{\gamma a y}(z) & \Delta_{a}(z)
		   				\end{array}\right) \, .
\eeq
In direct analogy with the standard treatment of the Schr\"odinger equation in quantum mechanics, we  have here introduced the `$z$-ordering' operator ${\mathcal T}_z$ in (\ref{eq:homsoln}).

In section \ref{sec:results} we will describe the results of numerically integrating equation (\ref{eq:EqofMotion}) for the inhomogeneous magnetic field discussed in section \ref{sec:Bsimul}. However, it is illuminating to first consider the simpler case of a homogeneous electron density and magnetic field in some domain of size $L$ (along the $z$-direction).
In this case,
the homogeneity makes the `$z$-ordering' and the  integral over $\dd z$ trivial. Furthermore, since only  photons with  polarisation parallel to  the magnetic field couple to axions, a simple rotation in the $x$-$y$ plane reduces the non-trivial part of the problem to that of a 2-body system of $\parphoton$ and $\axion$. The non-trivial part of the  $z$-evolution generator $\mathcal{M}$ can then be diagonalised by an orthogonal rotation by an angle $\theta$ satisfying
\beq
\tan\left( 2 \theta \right) = \frac{2 \Delta_{a \gamma}}{\Delta_a - \Delta_{\gamma}} \, ,
\eeq
where now $\Delta_{a\gamma} = B/2M$. The eigenvalues of $\mathcal{M}$ are given by $\lambda_{\pm} = \bar \lambda \pm \delta \lambda$, with $\bar \lambda = \frac{\Delta_a + \Delta_{\gamma}}{2}$ and
\beq
\delta \lambda = \frac{1}{2}\sqrt{(\Delta_a - \Delta_{\gamma})^2 + 4 \Delta_{a\gamma}^2} \, .
\eeq
The $z$-propagation is now trivial and, expressed  in the original basis,
results in the oscillation of an initially pure axion state, $\Ket{i} = (0,0,1)^{T}$, into the final state
\beq
\left(\begin{array}{c}
	\perpphoton\\
								\parphoton \\
	\axion \\
\end{array}\right)(L) =
\left(\begin{array}{c}
	0\\
								e^{- i( \vartheta +\frac{\pi}{2})} \sin(2 \theta) \sin( L \delta \lambda) \\
	e^{-i \vartheta}\left( \cos(L \delta \lambda) + i \sin(L \delta \lambda)\left(1-2\sin^2\theta \right) \right)\\
\end{array}\right) \, , \label{eq:outstate}
\eeq
where $\vartheta$ is a  phase that will be unimportant for our discussion. Thus, in a single domain with a homogeneous magnetic field,
the probability that an axion converts into a photon is given by %of either polarisation  is given by
\beq
\label{eq:convprob}
	P(a\rightarrow\gamma)=\sin^2(2\theta)\sin^2\left(L \delta \lambda\right)= \sin^2(2\theta)\sin^2\left(\frac{\Delta}{\cos 2 \theta}\right),
\eeq
where $\tan 2 \theta = \frac{2B_{\perp}\omega}{M m_{\rm eff}^2}$, $\Delta=\frac{m^2_{\rm eff}L}{4\omega}$ and $m_{\rm eff}^2=m_a^2 -\omega_{\rm pl}^2$.

For a single domain with a coherent magnetic field, the axion-photon conversion probability is completely determined by the angles $\theta$ and $\Delta$. For the values of electron density and magnetic field relevant for galaxy clusters, and for the values of $\omega$ and $M$ that we will consider,
the $\theta$ angle is always in the small-angle approximation,
\bea
	\theta & \approx & \frac{B_{\perp}\omega}{M m_{\rm eff}^2} = 8.1\ti 10^{-5}\left(\frac{n_0}{n_e}\right)\left(\frac{B_{\perp}}{1\mug}\right)\left(\frac{\omega}{200\hbox{\,eV}}\right)\left(\frac{10^{13}\hbox{\,GeV}}{M}\right) \, .
	\eea
Here, as in (\ref{eq:betamodel}), $n_0$ denotes the central electron density in the Coma cluster, $n_0 = 3.44\ti 10^{-3}\,$cm$^{-3}$.

By contrast, the $\Delta$ angle is not always small in clusters,
\bea
\Delta & = & 0.93\left(\frac{n_e}{n_0}\right)\left(\frac{200\hbox{\,eV}}{\omega}\right)\left(\frac{L}{1\kpc}\right) \, ,
\eea
and in section \ref{sec:results} we will find that much of the structure of CAB conversion in galaxy clusters can be understood in terms of the transition of  $\Delta$ between the small-angle and large-angle regimes.

Finally, for $\Delta \ll1$ and $\theta \ll1$,
the axion-photon conversion probability takes the simple form
\beq
\label{yoda}
	P(a\rightarrow\gamma) = 2.3\ti 10^{-8} \left(\frac{B_{\perp}}{1\mug}\frac{L}{1\kpc}\frac{10^{13}\hbox{\,GeV}}{M}\right)^2 \, .
\eeq
While equations (\ref{eq:convprob}) and (\ref{yoda}) are not directly applicable to axion-photon conversion in clusters,
%to the problem at hand,
we will still find them very useful
for understanding the general qualitative properties of axion-photon conversion.

%%%%%%%%%%%%%%%%%%%%%%%%%%%%%%%%%%%%%%%%%%%%%%%%%%%%%%%%%%%

\subsubsection{Axion-photon propagation II: Inhomogeneous magnetic fields}

The axion-photon conversion probabilities are computed by numerically simulating the propagation of an axion through the discretised magnetic field model discussed in section \ref{sec:Bsimul}. Since the lattice spacing of $0.5\kpc$ is much smaller than the cluster radius $r_c=291\kpc$, the electron density is slowly varying over each zone of constant magnetic field and may consistently be approximated as constant within each lattice zone. Thus within each zone the unitary `$z$-evolution' matrix is constant, and equation (\ref{eq:homsoln}) can be solved recursively from the $n$th lattice point to the next as
\beq
\left(\begin{array}{c}
	\Ket{\gamma_x} \\
	\Ket{\gamma_y} \\
	\axion \\
\end{array}\right)_{n+1} =  \exp \left(-i \omega s \mathbb{I} \ -i \left(\begin{array}{c c c}
							\Delta_{\gamma,\ n} & 0 & \Delta_{\gamma a x,\ n} \\
							0 & \Delta_{\gamma,\ n} & \Delta_{\gamma a y,\ n} \\
							\Delta_{\gamma a x,\ n} & \Delta_{\gamma a y,\ n} & \Delta_{a,\ n}
		   				\end{array}\right) s\right) \left(\begin{array}{c}
				\Ket{\gamma_x}\\
				\Ket{\gamma_y} \\
				\axion
		      \end{array}\right)_n \, , \label{eq:prop}
\eeq
where we have again denoted the lattice spacing by $s$. This way, an initial pure axion state will develop non-vanishing photon components as the particle propagates through the cluster.

The solution to equation (\ref{eq:prop}) is obtained, just as in the single-domain case, by first rotating to a basis in which the magnetic field is aligned with one of the coordinate axes and then diagonalising the non-trivial part of the $z$-propagation generator. This way, the axion-photon propagation can be solved exactly for each lattice point.

The propagation of the full 3-body system through the lattice is then achieved recursively by diagonalising, propagating the new fields to the next grid point, and finally rotating back to obtain the state with respect to the original reference basis.
Thus the state at the $(n+1)$th step is given by
\beq
	\Ket{n+1}= U_{1,n}^{\rm T}U_{2,n}^{\rm T}\mathcal{M}_{n}U_{2,n}U_{1,n} \Ket{n} \, ,
\eeq
where $U_{1,n}$ denotes the rotation required to align a coordinate axis with the local magnetic field direction, and $U_{2,n}$ denotes the diagonalisation of the unitary $z$-evolution operator at the $n$th step.

The probability of the axion converting into a photon is then computed as the sum of the squares of the $\Ket{\gamma_x}$ and $\Ket{\gamma_y}$ components of the final state. This procedure is done for each of the $2000^2$ grid points in the $x$-$y$ plane for 14 energies in the $25\,\mathrm{eV} - 2\keV$ range and for a vanishing axion mass.
By setting the axion mass to zero we make the approximation that $m_{\rm eff}^2$ is dominated by the plasma frequency. The plasma frequency is given as
\be
	\omega_{\rm pl} = 1.2 \times 10^{-12}\sqrt{\frac{n_e}{10^{-3}\hbox{cm}^{-3}}}\hbox{\,eV}
\ee
and is never much less than $\approx 10^{-12}\hbox{\,eV}$ in the system we are considering.
Axion masses $m_a \leq 10^{-13}\hbox{\,eV}$ therefore behave equivalently to a vanishing axion mass.
For axion masses $m \gg 10^{-12}\hbox{\,eV}$ (including the case of a QCD axion), $\theta \sim 1/m_a^2$ and $\Delta \sim m_a^2$, so that it is reasonable to expect that the conversion probabilities become suppressed relative to those we have obtained by a factor of $(10^{-13}\,{\rm eV} /m_a)^4$. For an axion mass $m \approx 10^{-12}\hbox{\,eV}$, detailed simulation would be required to study the resulting morphology and how it differs from the zero-mass case.

%%%%%%%%%%%%%%%%%%%%%%%%%%%%%%%%%%%%%%%%%%%%%%%%%%%%%%%%%%%

\subsubsection{Analysing the simulation data}

The simulation generates a $2000^2$ grid of $a \to \gamma$ conversion probabilities, each representing the probability of a single axion at energy $\omega$,
traversing Coma through a unit of cross-sectional area $(0.5\kpc)^2$, converting into a photon of the same energy.
We need to convert these probabilities into intrinsic source luminosities.  A redshift-distance converter is in \cite{Wright2006}.
Coma is at a well-determined redshift of $z=0.023$. We use the parameters from the magnetic field model of \cite{10020594}, with
$H_0 = 71\hbox{\,km\,s}^{-1}$ corresponding to $0.460\kpc$ per arcsecond and a luminosity
distance of $98.9\Mpc$. We note the soft excess analysis of
\cite{astroph0205473} assumed a Hubble constant of $H_0 = 75\hbox{\,km\,s}^{-1}$, corresponding to an angular scale
of $0.434\kpc$ per arcsecond and a luminosity distance to Coma of $93.6\Mpc$. These differences are small enough to be neglected compared to the other
statistical and systematic uncertainties in the extraction of the soft excess.

The overall CAB energy density is $ \rho_{\rm CAB} = \Delta N_{\rm eff} \frac{7}{8} \left( \frac{4}{11} \right)^{4/3} \rho_{\rm CMB}$.
Associated with this is a CAB number density $\dd N_a/\dd E$ set by the spectral shape, such that
$$
\int \dd E E \frac{\dd N_a}{dE} = \rho_{\rm CAB} \, .
$$
In terms of these the intrinsic excess luminosity associated with axion-photon conversion is given by
\be
\mc{L}_{\rm excess} = D_{\rm Coma}^2 \int \dd \Omega \dd E \, E \frac{\dd N_a}{\dd E} c P_{a \to \gamma}(\Omega, E)\, ,
\ee
where $D_{\rm Coma}$ is the physical distance to Coma and $\dd \Omega$ is a solid angle element (measured in $\hbox{arcmin}^2$).
We will restrict the energy integral to the $0.2 - 0.4\keV$ range, in accordance with \cite{astroph0205473}.

%%%%%%%%%%%%%%%%%%%%%%%%%%%%%%%%%%%%%%%%%%%%%%%%%%%%%%%%%%%
%%%%%%%%%%%%%%%%%%%%%%%%%%%%%%%%%%%%%%%%%%%%%%%%%%%%%%%%%%%

\subsection{Results} \label{sec:results}
In this section we discuss the results of the numerical simulation of the conversion probabilities, and we present for the first time a detailed description of the predictions of the CAB conversion scenario for the cluster soft excess.
\begin{table}
	\centering
	\begin{tabular}{| c | p{2.5cm} | p{2.5cm} | p{2.5cm} |} \hline
		& Model 1 & Model 2 & Model 3 \\
		& & & \\ \hline
		$\Lambda_{\mini}$ & $2\kpc$ & $2\kpc$ & $2\kpc$\\
		$\Lambda_{\maxi}$ & $34\kpc$& $5\kpc$&$100\kpc$ \\
		$n$ & $17/3$ &$17/3$ &$4$ \\
		$B_0$ & $3.9-5.4\mug$&$5.4\mug$ &$ 5.4\mug$ \\
		$\eta$ & $0.4-0.7$& $0.7$&$0.7$ \\ \hline
	\end{tabular}
	\caption{The  parameter values for the three simulations used. The first simulation (Model 1) is a Kolomogorov spectrum that fits the Faraday rotation data.
The second simulation (Model 2) is designed to show the effect of concentrating all power on small scales but is not a fit to Faraday rotation data. The third
simulation (Model 3) is a flat spectrum (in $k$-space) that fits Faraday rotation data and has most power on small scales.}
\label{tab:params}
\end{table}

\subsubsection{General features of axion-photon conversion}
\label{sec:general}
\begin{figure}[t]
	\subfloat[25\,eV]{\includegraphics[width=0.5\textwidth]{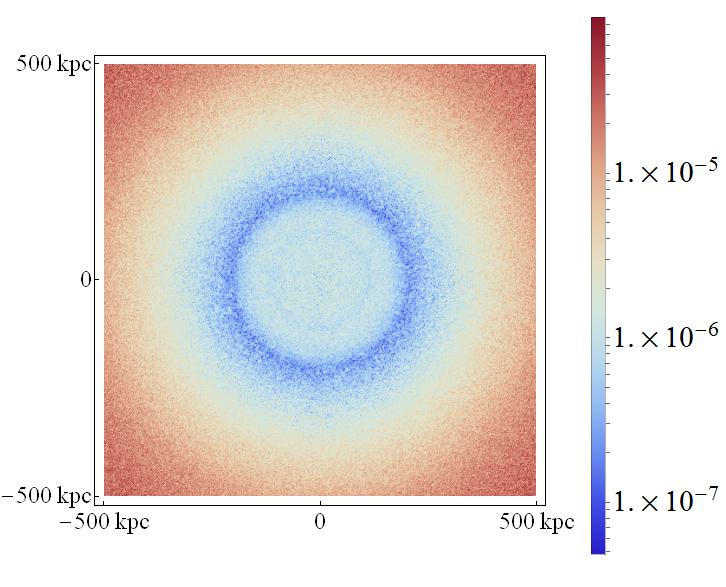} \label{fig:25eV}}
	\subfloat[50\,eV]{\includegraphics[width=0.5\textwidth]{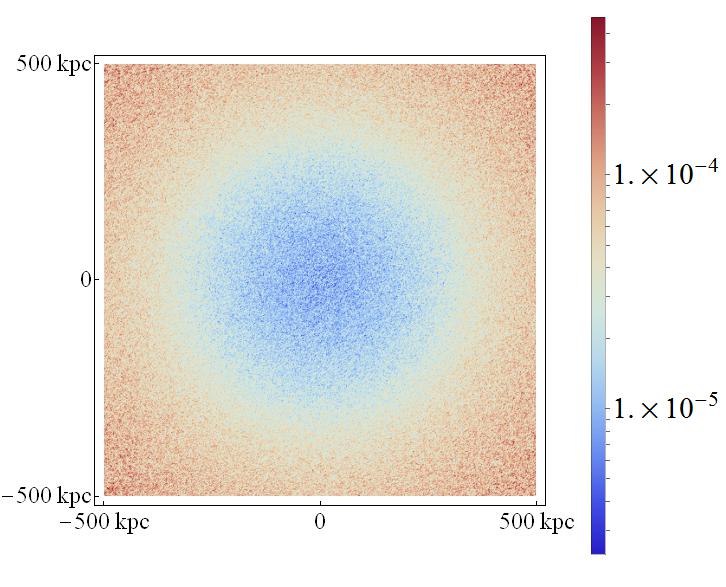}}\\
	\subfloat[75\,eV]{\includegraphics[width=0.5\textwidth]{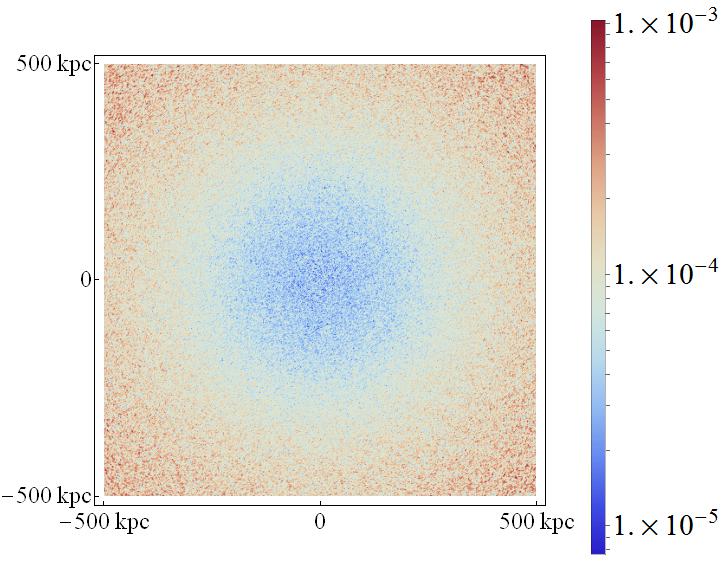}}
	\subfloat[100\,eV]{\includegraphics[width=0.5\textwidth]{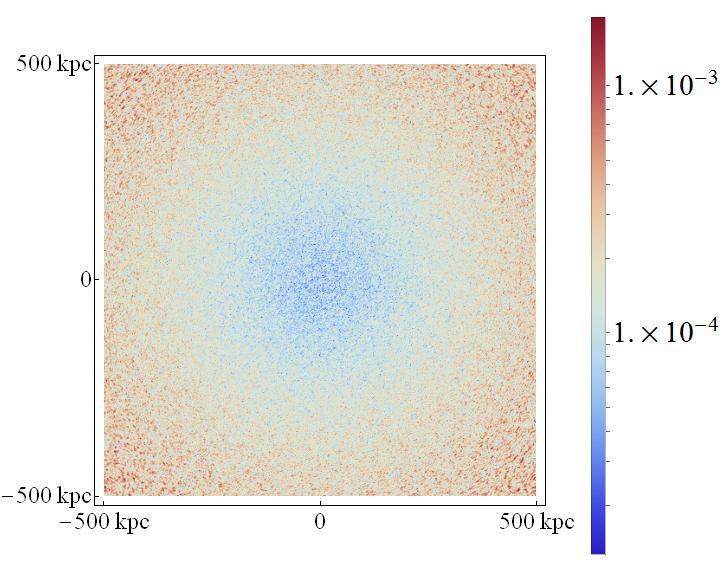}}
	\caption{Conversion probabilities for energies $25\hbox{\,eV}$ to $100\hbox{\,eV}$ for Model 1, with $\eta= 0.5$, $B_0 =4.7\, \mu$G and $M=7\ti 10^{12}$ GeV.
	}
\label{fig:run1plotsa}
\end{figure}
\begin{figure}[t]
	\subfloat[150\,eV]{\includegraphics[width=0.5\textwidth]{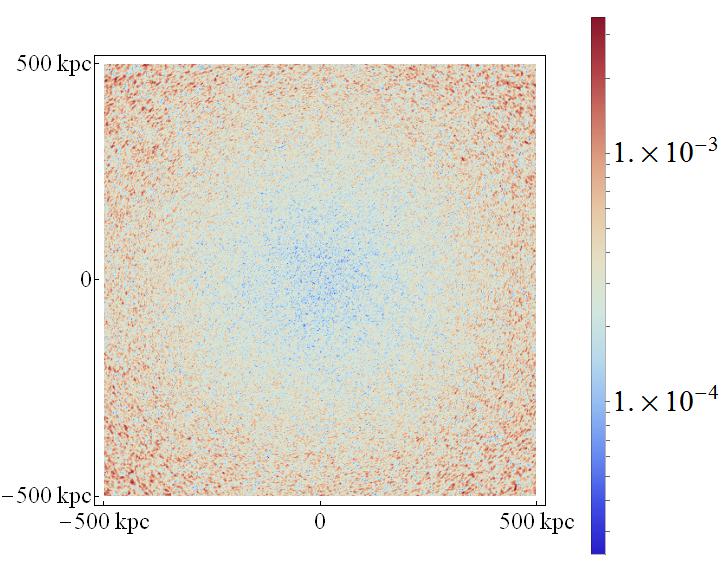}}
	\subfloat[200\,eV]{\includegraphics[width=0.5\textwidth]{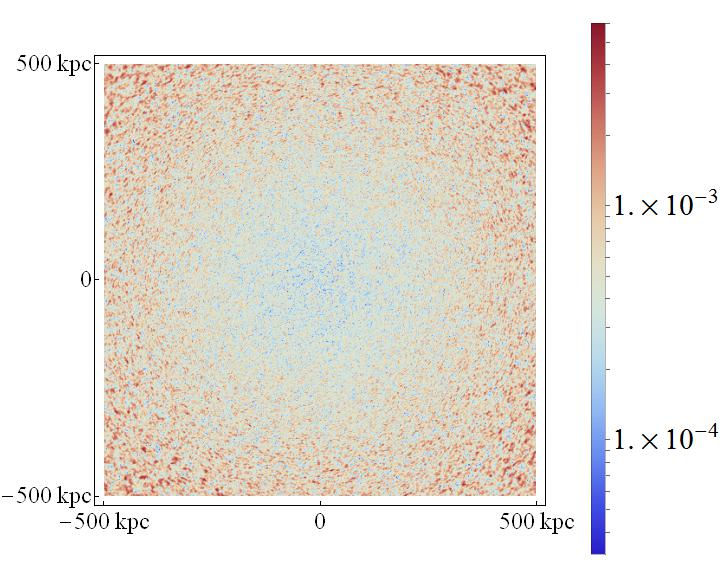}}\\
	\subfloat[300\,eV]{\includegraphics[width=0.5\textwidth]{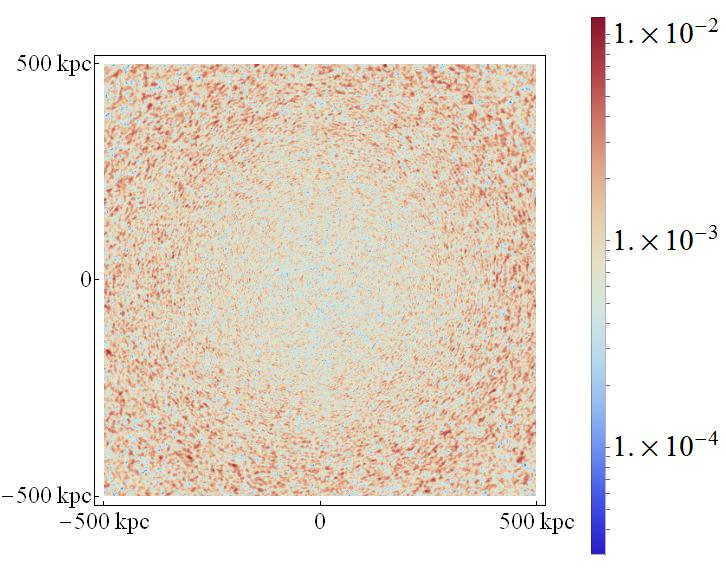}}
	\subfloat[400\,eV]{\includegraphics[width=0.5\textwidth]{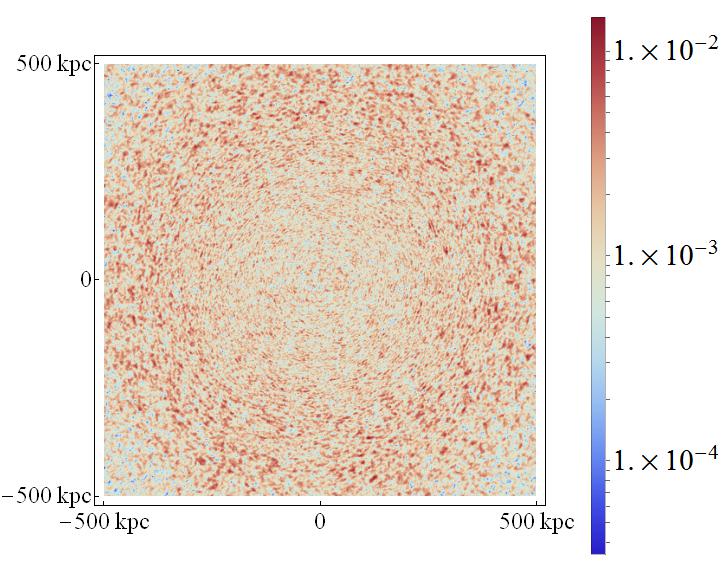} \label{fig:400eV}}
	\caption{Conversion probabilities for energies $150\hbox{\,eV}$ to $400\hbox{\,eV}$ for Model 1, with $\eta= 0.5$, $B_0 =4.7\, \mu$G and $M=7\ti 10^{12}$ GeV.}
\label{fig:run1plotsb}
\end{figure}
\begin{figure}[t]
	\subfloat[600\,eV]{\includegraphics[width=0.5\textwidth]{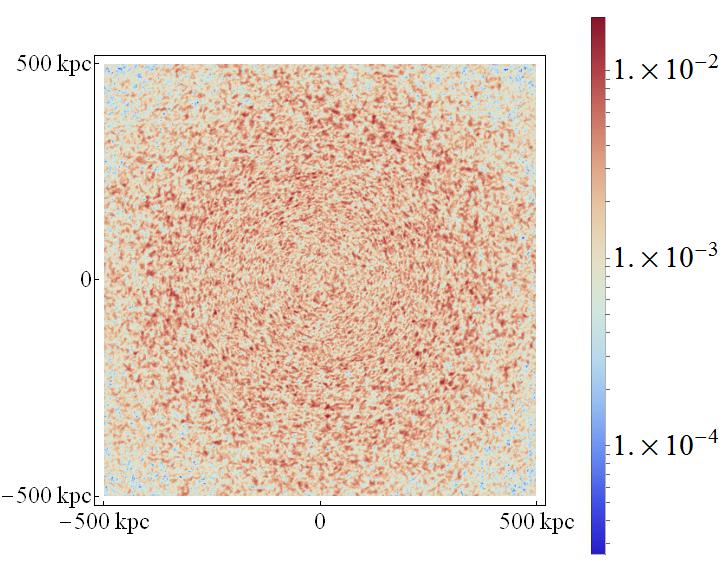}}
	\subfloat[800\,eV]{\includegraphics[width=0.5\textwidth]{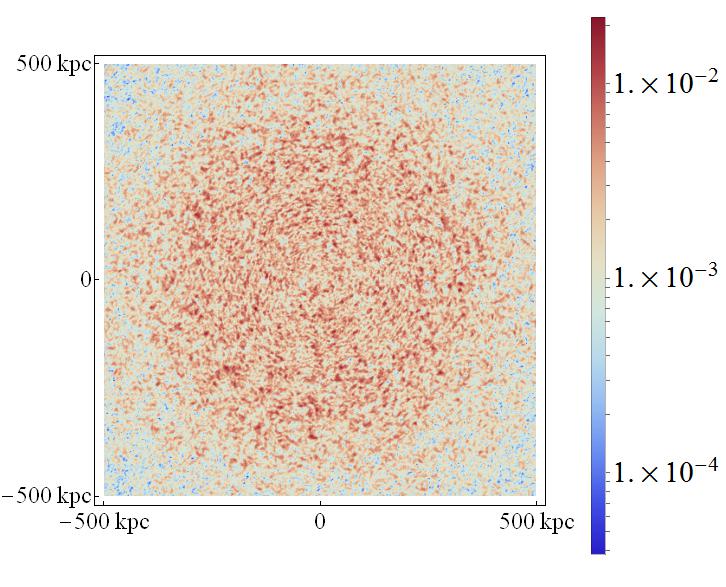}}\\
	\subfloat[1000\,eV]{\includegraphics[width=0.5\textwidth]{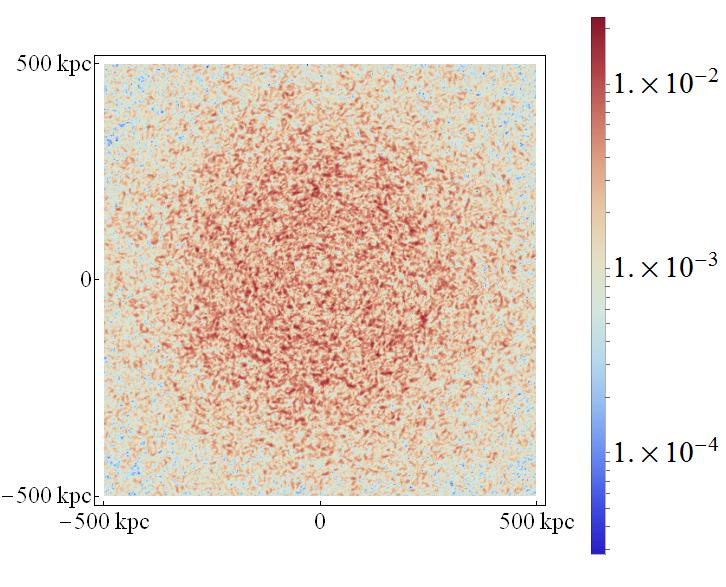}}
	\subfloat[2000\,eV]{\includegraphics[width=0.5\textwidth]{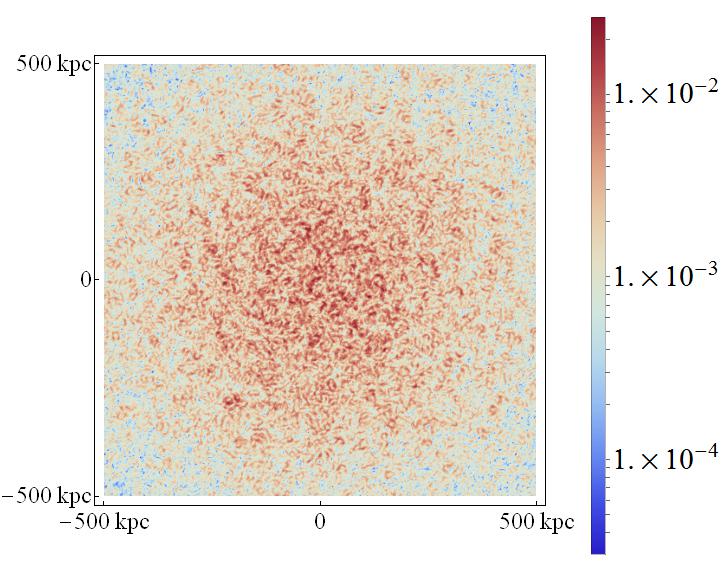}}
	\caption{Conversion probabilities for energies $600\hbox{\,eV}$ to $2\hbox{\,keV}$ for Model 1, with $\eta= 0.5$, $B_0 =4.7\, \mu$G and $M=7\ti 10^{12}$ GeV.}
\label{fig:run1plotsc}
\end{figure}

While several %of the detailed
properties of the simulated conversion probabilities and soft X-ray luminosities are sensitive to the detailed magnetic field model, there are also general features that are shared by all models we have considered. In this section we highlight these properties by using %the
 Model 1 of table \ref{tab:params} as our main example.

In this model, the stochastic magnetic field is generated with % This corresponds to
 a Kolmogorov power spectrum ($n=17/3$), with coherence lengths in the range  $2-34\kpc$. The best-fit values of the scaling of the total magnetic field with electron density, $\eta$, and the central value of the magnetic field, $B_0$, are then $\eta=0.5$ and $B_0 = 4.7\mug$, respectively.  In section \ref{sec:compare} we will also consider the effects of 1$\sigma$ variations of the parameters of the magnetic field model on the resulting conversion probabilities.

\begin{figure}
\centering
\includegraphics{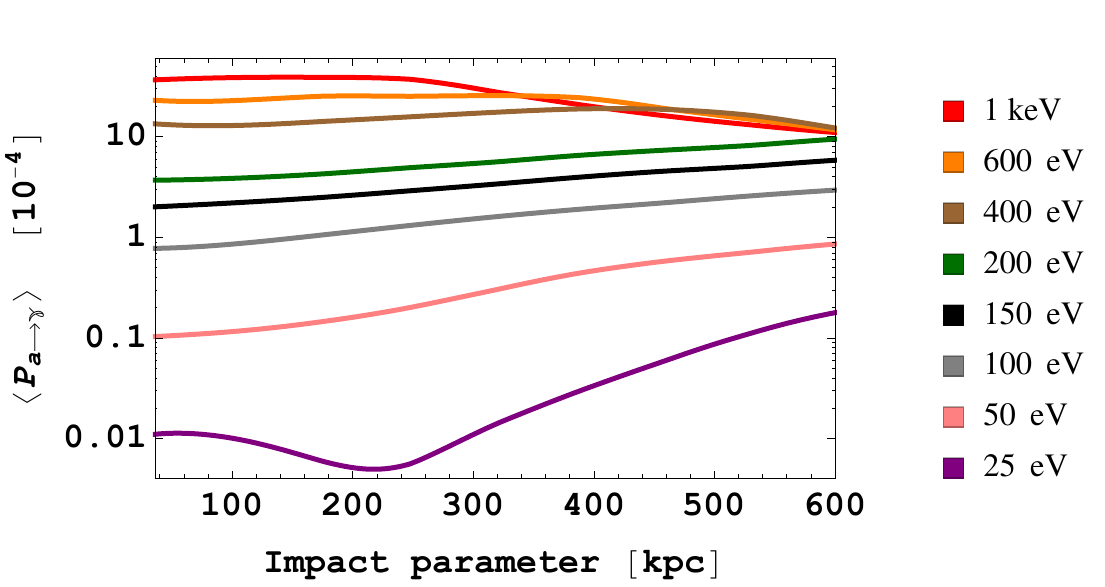}
\caption{Conversion probabilities as a function of impact parameter for Model 1 with $\eta=0.5$ and $B_0 = 4.7~\mu$G.}
\label{fig:radialModel1}
\end{figure}

The simulated conversion probability maps (best viewed in colour) for this model are shown in figures \ref{fig:run1plotsa}, \ref{fig:run1plotsb} and \ref{fig:run1plotsc} with a
pixel size of $(2\kpc)^2$. Figure \ref{fig:radialModel1} shows the conversion probabilities as a function of the impact parameter, for 8 energies from $25\keV$ to $2\keV$.

These figures illustrate two key features of the results that are ubiquitous in all
the magnetic field models we have studied.  First, the overall conversion probabilities increase with energy, up to a maximal energy at which they saturate.

Second, the morphology of the conversion probabilities is quite distinct at low and high energies. At low energies, the conversion probabilities are lowest for axions which pass through the
very centre of the cluster. As a function of increasing impact parameter, the conversion probabilities increase, reach a maximum at some intermediate radius, before again decreasing towards the edge of the cluster.  This behaviour is clearly visible in the $400\,\hbox{eV}$ plot (figure \ref{fig:400eV}) --- for lower energies than this the point of maximal conversion probability lies beyond the range of the simulation. At ultra-low energies (c.f.~the $25\,\hbox{eV}$ plot, figure \ref{fig:25eV}), a curious, unanticipated ring-like structure is visible. Here the conversion probabilities increase, decrease and then increase again. We discuss the  origin of this at greater length below.

In contrast, higher energy axions
 have a  maximal conversion probability at the centre of the cluster, with a monotonic decrease in conversion probability on going to larger radii. A crossover  between the high-energy regime  of `central dominance' and the low-energy regime  of `central deficit' can be observed for modes with energies of $400\,{\rm eV}< \omega<1\,{\rm keV}$ for Model 1.

In fact, both these generic features of the conversion probabilities can be understood from the
 single-domain solution of equation \ref{eq:convprob}, even though it is not fully applicable to the multi-scale fields
considered here.
The single domain conversion probability is (to leading order in $\theta$) given by
\be
\label{bcd}
P(a \to \gamma) \propto B^2_{0} \frac{\tilde \omega^2}{M^2} \tilde n(r)^{2(\eta-1)} \sin^2\left(0.93 \frac{\tilde L \tilde n(r)}{\tilde \omega}  \right) \, ,
\ee
where $\tilde n(r) = \frac{n_e(r)}{n_0}$, $\tilde L= \frac{L}{(1\kpc)}$ and $\tilde \omega = \frac{\omega}{(200\,{\rm eV})}$. Here we have factored out the dependence of the total magnetic field on the electron density, as in equation (\ref{eq:Btot}). The fractional electron density $\tilde n(r)$ is completely determined by the $\beta$-model, c.f.~equation (\ref{eq:betamodel}), and decreases from unity at the cluster centre to $\approx 0.15$ at $r=600\kpc$.

We now note that for either sufficiently large $\omega$ or sufficiently small $\tilde n(r)$, the argument of the $\sin$ function becomes small.
In the small-$\Delta$ approximation the conversion probabilities are given by
 \bea
\label{largeomega}
P(a \to \gamma)_{\textrm{single domain}} & \propto & \frac{B_0^2 L^2}{M^2} \tilde n_e^{\eta}(r) \, , \\
\label{largeomega2}
P(a \to \gamma)_{\textrm{per unit length}} & \propto & \frac{B_0^2 L}{M^2}  \tilde n_e^{\eta}(r)\, .
\eea
Thus, according to the single-domain formula, sufficiently far away from the cluster centre the small-$\theta$ \emph{and} small-$\Delta$ approximations should be valid for all energies above a certain cut-off.
Evidently, at large radii the small-angle approximation appears as an `attractor' with a radial dependence completely determined by the modulation of the magnetic field with $\tilde n(r)^{\eta}$.

As the impact parameter is decreased and $\tilde{n}(r)$ increases, some modes will cease to be well described by the small-$\Delta$ approximation and will rather require the full equation (\ref{bcd}). Such modes will leave the small-angle `attractor' solution.   According to the single-domain formula, modes with sufficiently low energies may undergo several $2\pi$ rotations of $\Delta$ as the impact parameter is decreased towards the centre of the cluster, and these modes will in particular exhibit rings of decreased probabilities as $\Delta$ comes close to an integer multiple of $\pi$.

However, for an axion traversing multiple magnetic field domains with slightly varying electron densities and coherence lengths,  some of the detailed features of the single-domain probabilities can be expected to be `washed out'.  In particular, in the large $\Delta$ regime it is reasonable to
approximate $\langle \sin^2 \Delta \rangle = \half$ for axions traversing multiple domains of slightly varying size. Then for $B \propto B_0 n_e^{\eta}$, we have
\bea
\label{abcd}
P(a \to \gamma)_{\textrm{single domain}} & \propto & \frac{B_0^2 \tilde \omega^2}{M^2} \tilde n_e^{2(\eta-1)} \, , \\
\label{abcd2}
P(a \to \gamma)_{\textrm{per unit length}} & \propto & \frac{B_0^2 \tilde \omega^2}{L M^2} \tilde n_e^{2(\eta-1)} \, .
\eea
As in all cases we consider $\eta<1$, this gives increasing conversion probabilities  as the electron density decreases with radii. This increase will continue until the small-$\Delta$ regime is reached, when the conversion
probability is again described by equation (\ref{largeomega}).

\begin{figure}
\centering
\subfloat[Model 1]{\includegraphics[width=0.5\textwidth]{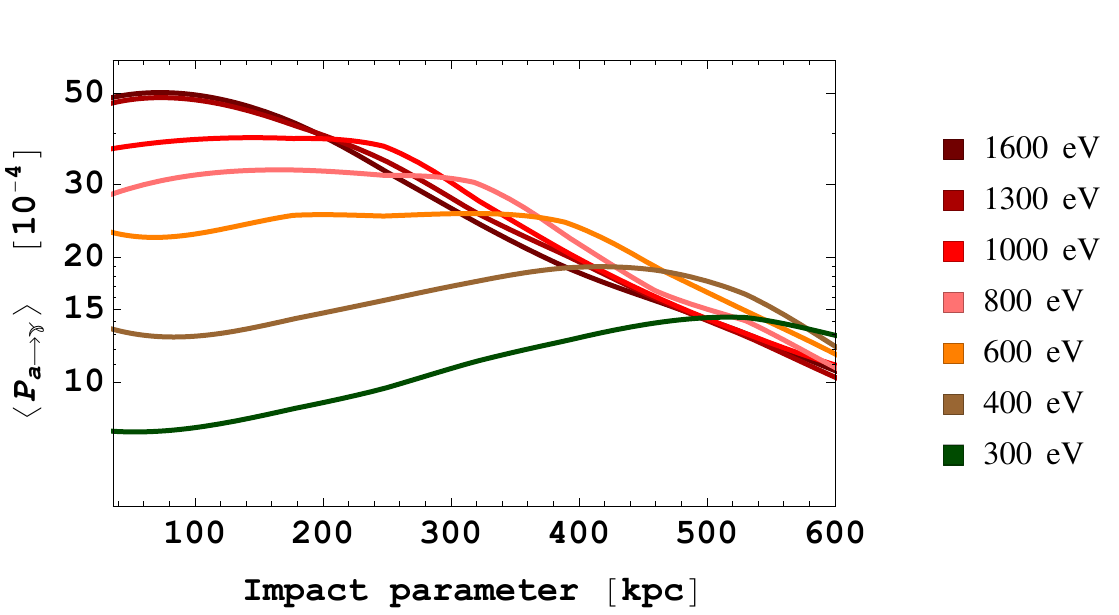} \label{fig:radialModel1b}}
\subfloat[Model 2]{\includegraphics[width=0.5\textwidth]{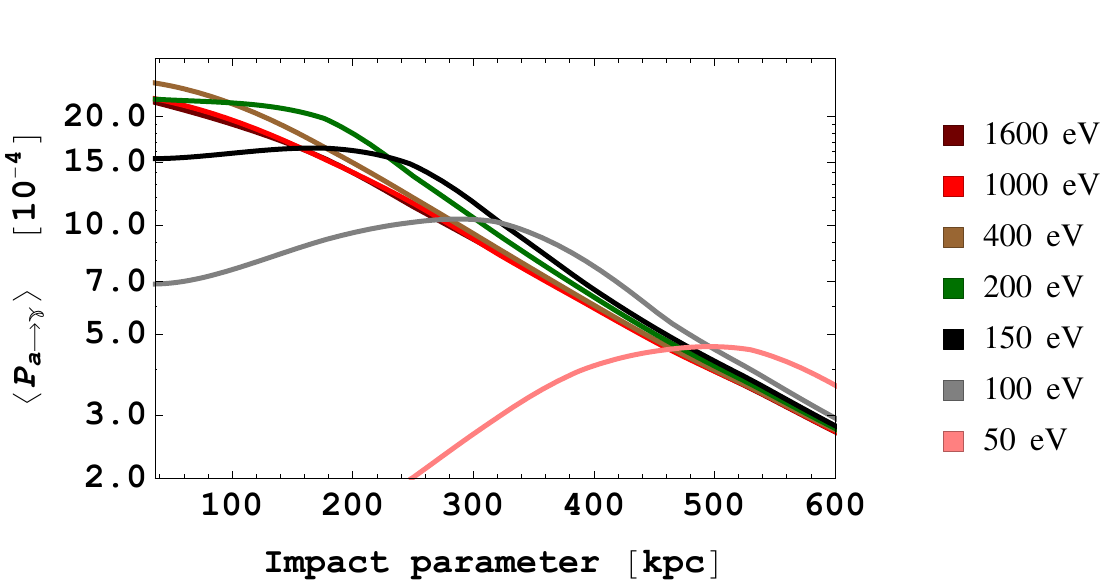} \label{fig:radialModel2}}
\caption{At large impact parameter the conversion probabilities tend to the small angle approximation, as here illustrated for Model 1 and Model 2. }
\end{figure}

We now note that several of the  features predicted from the single-domain formula also appear in the radial probabilities emerging from the
full numerical simulation with multi-scale magnetic fields, c.f.~figures \ref{fig:radialModel1} and \ref{fig:radialModel1b}. From these figures, we note that modes with $\omega \gtrsim 400\,\hbox{eV}$ have equal conversion probabilities at large radii (small $\tilde n(r)$), consistent with these modes entering the small-$\Delta$ approximation.  At smaller impact parameter (corresponding to larger maximal $\tilde n(r)$), the lower energy modes begin to decouple from the small-$\Delta$ approximation with ultimately decreasing probabilities as a result, perfectly consistent with the single-domain result. Only the highest energy modes stay in the attractor curve
 for any impact parameter,  as illustrated in figure \ref{fig:radialModel1b} by the modes with $\omega = 1300\,\hbox{eV}$ and $\omega= 1600\,\hbox{eV}$.

 The single domain analysis  also predicts the presence of regions with highly suppressed conversion probabilities as $\Delta$ approaches an integer multiple of $\pi$, and we note that this feature  most likely explains the  ring structure of the conversion probability in the $25\,\hbox{eV}$ plot.

We can use the qualitative consistency of the conversion probabilities with the single-domain result to
generate a heuristic estimate of an `effective coherence length' of the magnetic field.   For the high-energy modes in figure \ref{fig:radialModel1b} we have argued that $\Delta \ll1$ at large impact parameter, and we may heuristically associate the radius of maximum conversion probability, $r_{\maxi}(\omega)$, for each mode with the phase $\Delta = \pi/2$ in the single-domain formula.  By furthermore noting that the largest contribution to the conversion probability for modes close to the small-$\Delta$ approximation will come from the region closest to the cluster centre, we have
\beq
\frac{\pi}{2} = 0.93 \frac{\tilde L \tilde n(r_c(\omega))}{\tilde \omega}  \, . \label{eq:effL}
\eeq
From this formula, we can extract an `effective coherence length' $\tilde L$ for each mode that has a peaking conversion probability within the range studied. For Model 1 this range corresponds to modes with energies $300\,{\rm eV} \leq \omega \leq 1\,{\rm keV}$, as indicated in figure \ref{fig:radialModel1b}. The decoupling of all modes is consistent with the single-domain estimate for effective coherence lengths in the $13-15\kpc$ range. In comparison, the full numerical simulation involves magnetic fields coherent over all scales from $2\kpc$ to $34\kpc$, with a mean coherence length of $\approx 10\kpc$.

The single-domain intuition also holds for other magnetic field models. From figure \ref{fig:radialModel2}, we note that Model 2  may be associated with an `effective coherence length' of the magnetic field %from equation (\ref{eq:effL})
in the \hbox{$2.0-2.2\kpc$} range, based on the peak positions of modes with $50\,{\rm eV}< \omega<200\,{\rm eV}$. The full multi-scale model has  coherence lengths  in the $2-5\kpc$ range, with a mean value of $3.2\kpc$.  These estimates indicate that the physical picture of axion-photon conversion motivated by the single-domain analysis is also quite accurate for more complicated magnetic field configurations such as the multi-scale configurations considered in this paper.

\begin{figure}[t]
\centering
\includegraphics[width=0.85\textwidth]{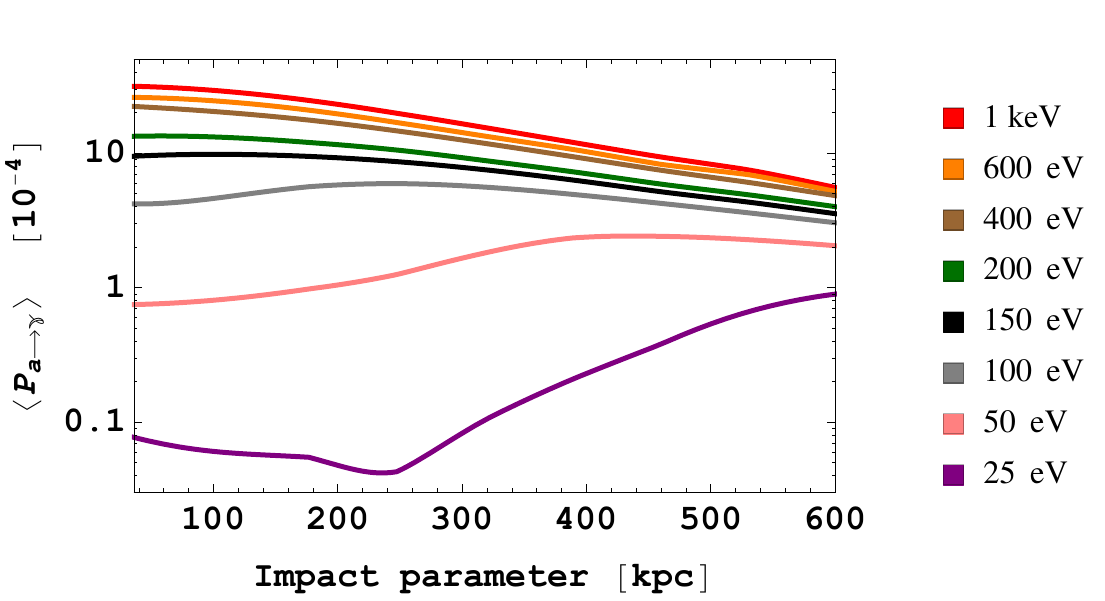}
\caption{Mean conversion probability as a function of radius from the centre of Coma, for Model 3 of table~\ref{tab:params} and with $M= 5.7 \ti 10^{12}\hbox{\,GeV}$. }
\label{fig:radialModel3}
\end{figure}

Model 3 is distinguished by having the largest range of scales in the magnetic field, from \hbox{$2 - 100\kpc$}. As we see in figure \ref{fig:radialModel3},
for this model at large radius ($500-600\kpc$) the conversion probabilities have not converged to a small-angle approximation.
We can again understand this behaviour using the single-domain formula.

The greater range of coherence lengths imply that even
axions traversing the cluster at large impact parameter are likely to encounter domains in which
the small-$\Delta$ approximation is not valid. Increasing $\omega$ always has the effect of decreasing $\Delta$ and thereby bringing a larger fraction of the traversed distance
into
the small-angle approximation, which results in an increased overall conversion probability according to equations (\ref{largeomega2}) and (\ref{abcd2}).  This explains
the increase in conversion probabilities with $\omega$ in Model 3, even at the largest radii.

A further difference in figure \ref{fig:radialModel3} compared to figures \ref{fig:radialModel1b} or \ref{fig:radialModel2} can also be understood.
For model 3, at large radii the conversion probabilities are roughly similar for the higher energy modes, with consistent small increases as you go to higher energy.
This behaviour is absent for models 1 and 2. We can understand this through the different power spectra of the models. For the Kolmogorov spectra of models 1 and 2, power dominantly
lies in the largest coherence lengths. The conversion probabilities are then highly suppressed for axion modes that have insufficient energy to reach the small-$\Delta$ regime, given this range of coherence lengths.  For the truncated spectrum of model 2 and the flat spectrum of model 3, more power of the magnetic field is allocated to shorter scales, and even lower energy modes can reach the small-$\Delta$ approximation. Owing to the wide range of coherence length in the Model 3 magnetic field, axions traversing the cluster will still pass through regions which are not well-described by a small-$\Delta$ approximation --- even for relatively high energy modes. As the axion mode energy is increased, only a small additional fraction of the large coherence lengths are brought into the small-$\Delta$ approximation, which explains the gradual approach to the small-$\Delta$ approximation in this case.

Let us finally for the sake of clarity remark on features of the conversion probabilities that are not well-captured by the single-domain formula. %, equation (\ref{bcd}).
As $\Delta$ approaches $\pi/2$, the single-domain formula predicts a relative decrease in the conversion probability with respect to the small-$\Delta$ approximation, yet the full numerical simulations exhibit conversion probabilities with a clear trend of `overshooting' the small-angle attractor at this point. %\dm{Can we explain this?}

A final general comment about the simulations: it is useful to know how much variation one can expect purely from repeating a simulation with identical magnetic field parameters.
By repeating simulations with the Model 1 parameters, we found that the averaged conversion probabilities within each annulus varied by at most $9\%$, where in most cases the difference was less than $5\%$. We note that the most significant variations occurred for larger energies. We thus conclude that our magnetic field model does not generate large fluctuations in conversion probabilities. However, to account for this when we plot comparisons between simulated and observed luminosities, we include a statistical error of 10\% on our values.

%%%%%%%%%%%%%%%%%%%%%%%%%%%%%%%%%%%%%%%%%%%%%%%%%%%%%%%%%%%

\subsubsection{Comparison with Observed Luminosities} \label{sec:compare}
We may now compare the predictions of axion-photon conversion of a CAB in the Coma cluster to the actual observations of the soft excess with ROSAT, based on the analysis of Bonamente et al.~\cite{astroph0205473}. Specifically, we will focus on  the %This work quotes results as an
overall unabsorbed excess luminosity in the $0.2 - 0.4\keV$ band for various different annular regions around the centre of Coma. % to the corresponding prediction from CAB conversion.

Since the spectral shape of the soft excess is poorly known, the analysis of \cite{astroph0205473} quotes results for two different spectral fits to the excess emission:
the first employs a power-law spectrum with photon index $1.75$ (so that the excess flux is $\dd N_{\gamma}/\dd E \sim \nu^{-1.75}$),  and the second is based on a thermal spectrum with $T = 80\,$eV.
These  results --- which differ from each other by an overall factor of $\approx 2.4$ --- are shown in table \ref{tab:excess}.
\begin{table}
\centering
\begin{tabular}{| c | c | c | c | %c |
}
\hline
Region (arcminute) & %Area ($\pi $ arcmin$^2$) &
 $L_{NT} (10^{41} \hbox{erg s}^{-1})$ & $L_{thermal}  (10^{41} \hbox{erg s}^{-1})$ \\
& & %&
\\
\hline
0 - 3 & %9 &
 11 & 4.6 \\
3 - 6 &% 27 &
 22 & 9.1 \\
6 - 9 & %36 &
 26 & 10 \\
9 - 12 &% 63 &
 25 & 10 \\
12 - 18 &% 180 &
 47 & 21 \\
\hline
\end{tabular}
\caption{The results of \cite{astroph0205473} for excess luminosity from the Coma cluster.}
\label{tab:excess}
\end{table}
While neither of these spectral models  are in exact agreement with the shape of the photon spectrum obtained from CAB conversion (which we will discuss in more detail in section~\ref{sec:constraints}), the thermal bremsstrahlung spectrum has an exponentially decreasing tail in the $0.2-0.4\keV$ range, as does the CAB spectrum for mean axion energies $\langle E_{\rm CAB} \rangle \lesssim 200\,$eV.   We will therefore use the thermal excess luminosities of table \ref{tab:excess} when comparing the predictions of our model to the data. We also note that the model uncertainty in the extraction of the soft excess mostly affects the overall luminosity and to a much smaller degree its spatial distribution. In our model,  the overall luminosity has a simple dependence on the values of $M$ and $\Delta N_{\rm eff}$, and the uncertainty in the overall luminosity translates into an uncertainty in $\Delta N_{\rm eff}/M^2$.

We will now present our main results for the CAB explanation of the soft excess in Coma. We will first discuss luminosities from axion-photon conversion in the Model 1 magnetic field (including 1$\sigma$ variations of the model parameters $\eta$ and $B_0$)  and we will then turn to Model 2 and Model 3.

\begin{figure}[t]
	\centering
	\includegraphics[width=0.9\textwidth]{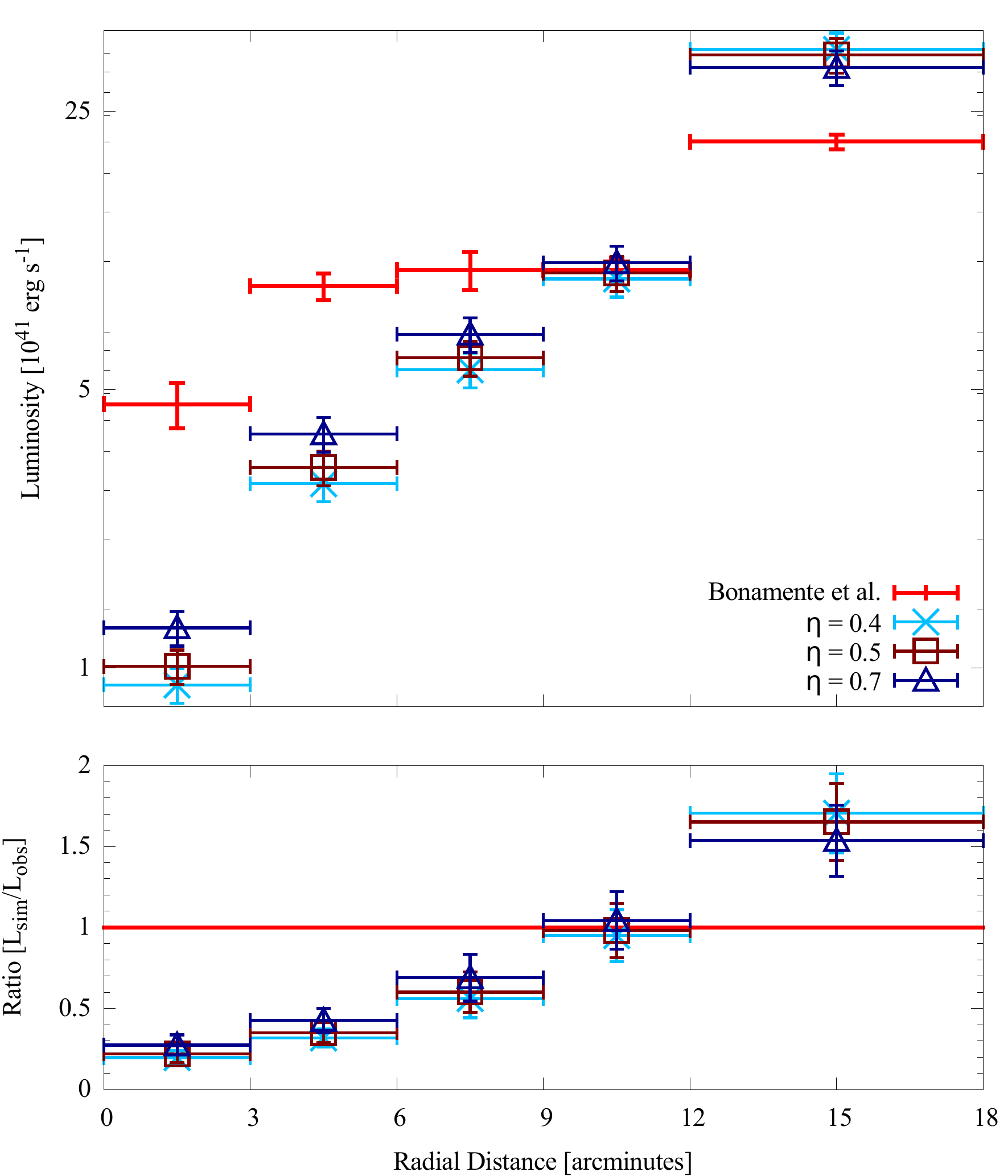}
	\caption{Luminosity comparison for Model 1 with different $\eta$ values, compared to the `thermal' excess data. For $\Delta N_{\rm eff} = 0.5$ and  $\langle E_{\rm CAB} \rangle = 150\,$eV, normalisation of the integrated luminosities give $M = 6.1\times 10^{12}$, $6.7\times 10^{12}$ and $6.5\times 10^{12}\hbox{\,GeV}$ for $\eta=0.4$, $0.5$ and $0.7$, respectively.}
\label{fig:etacomp}
\end{figure}

\paragraph{Model 1:}
Figure \ref{fig:etacomp} shows the comparison between  the observed (thermal) luminosity and that produced by the baseline model of \cite{10020594} and two other models related by  $1 \sigma$ variations of the model parameters.
The integrated luminosity in the $0.2- 0.4\keV$ range has been normalised to the total luminosity of the soft excess in the same range by varying $M$ independently for each model.  In all cases $\Delta N_{\rm eff} = 0.5$ has been assumed, but we emphasise that alternate values for these parameters that normalise luminosities can be obtained by scaling $\Delta N_{\rm eff} \rightarrow \lambda \Delta N_{\rm eff}$ and $M \rightarrow \sqrt{\lambda} M$.

From figure \ref{fig:etacomp}, we note that while the luminosity in each bin is within observations by a factor of a few, there is a clear tendency
 to underproduce photons in the centre and overproduce them in the outskirts. Axion-photon conversion in the Model 1 magnetic field therefore 
 does  not provide a particularly good description of the Coma soft excess.

The three variations of Model 1 in figure \ref{fig:etacomp} correspond to the magnetic field model parameters which best fit the Faraday rotation measures, here denoted by $\eta= 0.5$, and 1$\sigma$ variations to $\eta=0.4$ and $\eta=0.7$. In all cases, axion-photon conversion under- and over-produces photons in the inner and outer regions respectively. Understandably, increasing  $\eta$ so that the magnetic field falls off more rapidly with radius while simultaneously increasing $B_0$ to match Faraday rotation measures results in more luminosity to smaller radii relative to larger radii.
We note that these variations  are not large enough to make the CAB prediction of the morphology of the soft excess compatible with observations.

However, despite the poor fit of the axion-converted photon luminosities to the soft excess for Model 1, it would be premature to conclude that the conversion of the CAB cannot explain the soft excess. Even before considering systematic differences between the magnetic field models of \cite{10020594} and the actual magnetic field in Coma,
as discussed in section \ref{sec:ComaB}, Faraday rotation measures only constrain the magnetic field model up to degeneracies in the spectral index of the vector potential, $n$, and the Fourier mode cut-off scale $\Lambda_{\maxi}$. In Model 3, we consider a magnetic field model which provides an equally good fit to Faraday rotation measures as Model 1, but with $n=4$ and \hbox{$\Lambda_{\maxi} = 100\kpc$} (as opposed to $n=17/3$ and $\Lambda_{max} = 34\kpc$ for Model 1). As in equation (\ref{eq:Bspectrum}), the stochastically generated magnetic field prior to modulation by $n_e(r)$ has $\tilde B(k)_{(\rm gen.)} \sim k^{(-n+2)/2}$, so that for $n=4$ the power integral $\int \dd k k^2 \tilde B(k)^2_{(\rm gen.)}$ has constant support from $k_{\mini}$ to $k_{\maxi}$.  Equivalently, magnetic fields with $n>4$ locate more power to smaller $k$-numbers and larger physical scales.

Thus, a key difference between Model 1 and Model 3 is the distribution of effective coherence lengths. In order to highlight the effect of concentrating more  power of the magnetic field on smaller scales, we first consider a toy magnetic field model, which does not provide a good fit to Faraday rotation
measures.\footnote{Note that the small scales in this model are however not necessarily unphysical. Faraday rotation constrains the magnitude and coherence lengths of the parallel component of the magnetic field along the line of sight, whereas axion conversion involves the transverse components. The magnetic field models used here make these equal by assumption, but if the latter is actually smaller than the former by a factor of a few, this model could still be consistent with Faraday rotation.}
This is our Model 2, to which we now turn.

\begin{figure}[t]
	\centering
	\includegraphics[width=0.9\textwidth]{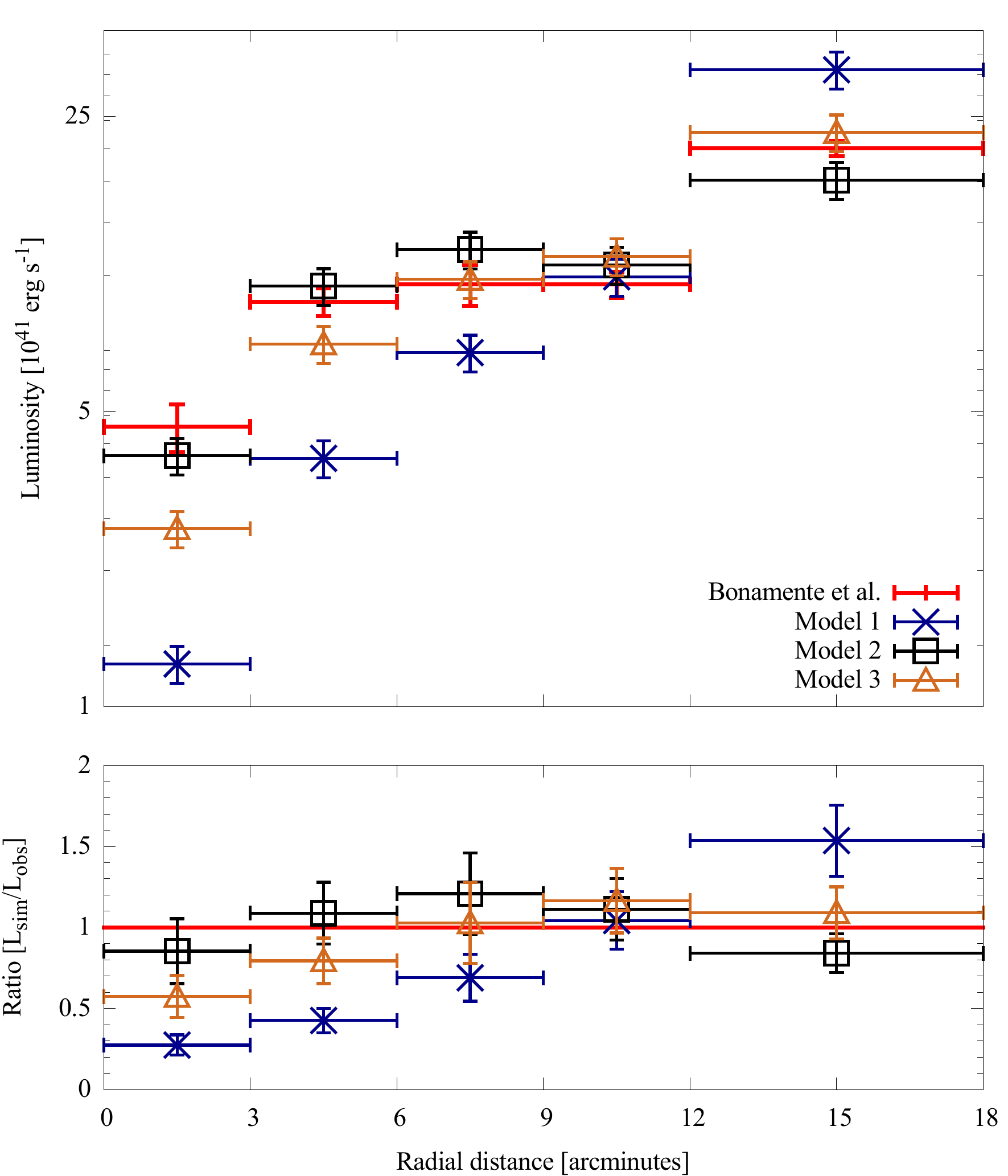}
	\caption{Luminosity comparison for the different models compared to the 'thermal' excess data. For $\Delta N_{\rm eff} = 0.5$ and  $\langle E_{\rm CAB} \rangle = 150\,$eV, normalisation of the integrated luminosities gives $M = 6.5\times 10^{12}$, $5.2\times 10^{12}$ and $5.7\times 10^{12}\hbox{\,GeV}$ for Models 1 ($\eta=0.7$), 2, and 3 respectively.}
\label{fig:modelcomp}
\end{figure}

\paragraph{Model 2:}
In this model
the generated magnetic field, prior to modulation by a function of the electron density, only varies on scales between $2-5\kpc$. In this range, the magnetic field varies with $n=17/3$ and the modulation with electron density is obtained with
$\eta=0.7$, setting $B_0 = 5.4\,\mug$. %We see from
The simulated photon luminosities from this model match the observational data for the soft excess very well, as is shown in figure \ref{fig:modelcomp}.

Based on our discussion in section \ref{sec:general} on the radial dependence of the simulated conversion probabilities, we may interpret the improved fit as due to a decreased `effective coherence length', resulting in modes approaching the small-$\Delta$ attractor at smaller radii, c.f.~figures \ref{fig:radialModel1b} and \ref{fig:radialModel2}.

\paragraph{Model 3:}
We now return to magnetic field models consistent with observations of Faraday rotation measures in Coma, but focus on models which concentrate more power on smaller scales relative to Model 1.  The effective degeneracy between values of $n$ and $\Lambda_{\maxi}$ allows for reducing $n$ by simultaneously increasing $\Lambda_{\maxi}$, as is illustrated in figure 16 of reference \cite{10020594}. Here again, we take $\eta=0.7$ and $B_0 = 5.4~\mu$G. The resulting photon luminosities from CAB conversion are shown in figure \ref{fig:modelcomp} and again show a good agreement with the observed soft excess.

The conclusions to draw from these are that an explanation of the soft excess via axion-photon conversion appears to require the transverse components of the magnetic field
to have more power on shorter scales than in the Kolmogorov spectra of \cite{10020594}. This can be achieved either by allowing a flatter spectrum, so as to be consistent with Faraday rotation measures even for a Gaussian magnetic field, or by having different coherence lengths for transverse and parallel components of the magnetic field.

%%%%%%%%%%%%%%%%%%%%%%%%%%%%%%%%%%%%%%%%%%%%%%%%%%%%%%%%%%%

%%%%%%%%%%%%%%%%%%%%%%%%%%%%%%%%%%%%%%%%%%%%%%%%%%%%%%%%%%%

\subsubsection{Constraints on the axion-photon coupling}
\label{sec:constraints}

\begin{figure}
\includegraphics[width=0.9\textwidth]{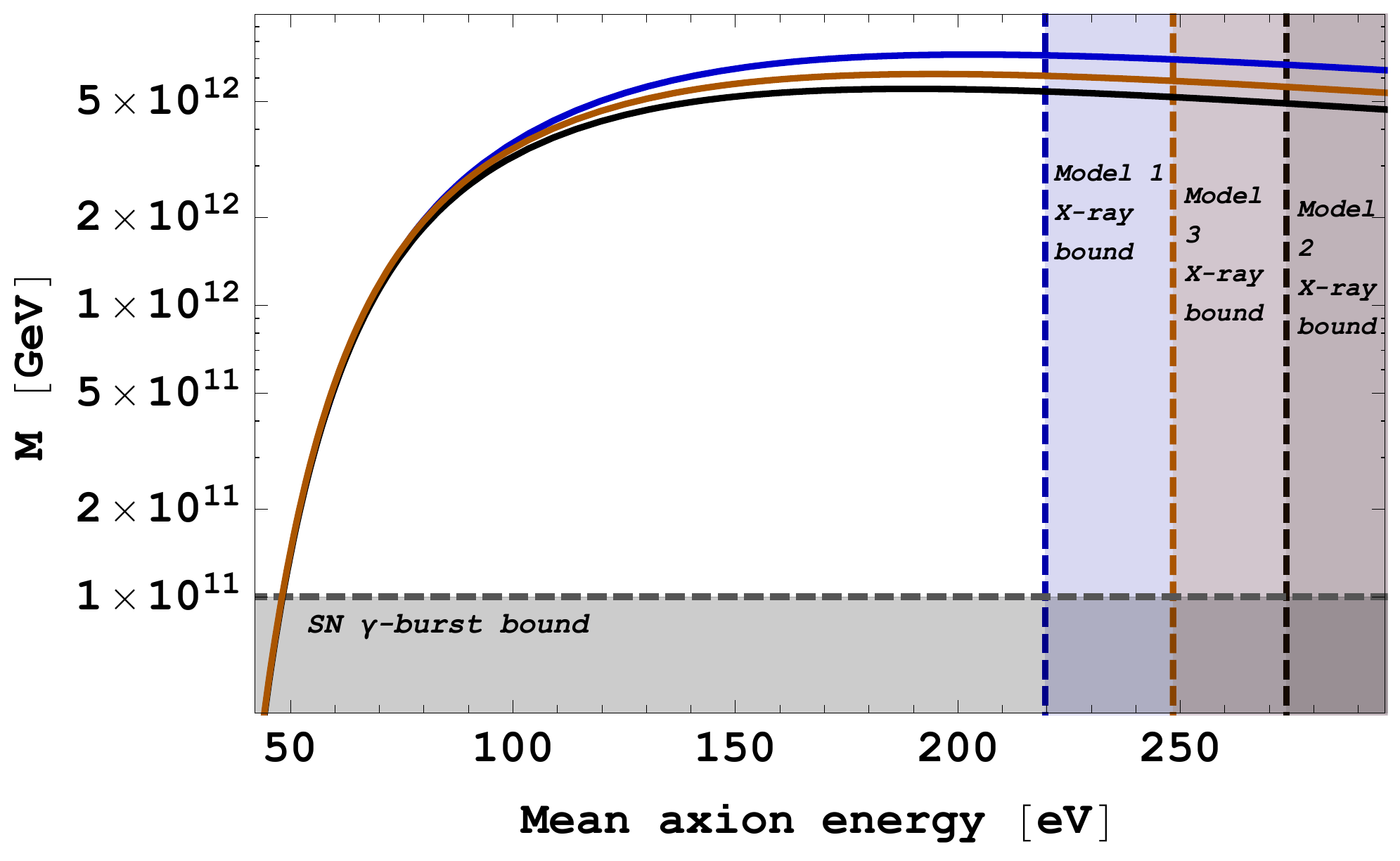}
\caption{The values of $M$ required to normalise the total soft excess from axion-photon conversion in the $0.2-0.4\keV$ band to the observed total excess luminosity with the central $18$ arcminutes of Coma as a function of $\langle E_{\rm CAB}\rangle$ for $\Delta N_{\rm eff} = 0.5$. Model 1 is represented by the blue solid curve, Model 2 by the black curve and Model 3 by the orange curve. The supernova $\gamma$-burst bound is indicated by a dashed grey line, and the bounds from overproduction of X-rays in the $0.5-0.6\keV$ range are indicated by a vertical dashed line for each model.    }
\label{fig:MvsE}
\end{figure}

Having established that the CAB explanation of the cluster soft excess is in reasonable agreement with observations for magnetic field models motivated by observations of Faraday rotation measures, we will now discuss how additional observational constraints give rise to a relatively limited range of possible values for $M$ and the mean CAB energy, $\langle E_{\rm CAB} \rangle$.

Strong upper bounds on the axion-photon coupling --- and thereby lower bounds on $M = g_{a\gamma\gamma}^{-1}$ --- have been obtained by laboratory experiments, helioscopes, and may also be inferred from astrophysical arguments (see e.g. \cite{12105081} for a recent review).  For light axions, the CAST search for solar axions has set a bound $M> 10^{10}\,$GeV. Proposed experiments looking either for light shining through a wall, such as ALPS-II,  or for solar axions, such as IAXO, are expected to improve this bound by a factor of $10-15$. An astrophysical bound based on the anomalous energy losses of horizontal branch stars due to axion emission similarly gives $M>10^{10}\,$GeV \cite{RaffeltBook1996}, with similar bounds also being attainable from `blue loop' Helium burning massive stars \cite{12101271}.  Moreover, the absence of $\gamma$-ray bursts in coincidence with neutrinos from Supernova 1987A  provides a bound of $M> 10^{11}\,$GeV for light axions ($m_a\lesssim10^{-9}\,$eV) \cite{Brockway:1996yr, Grifols:1996id}.

In addition to limits,
certain values of the axion parameters have also been suggested to be hinted by anomalous astrophysical processes. Spectra from active galactic nuclei (AGN) in $\gamma$-rays extend to the multi-TeV regime, even though scattering off ambient starlight at these energies should provide attenuation through production of $e^+e^-$ pairs. This apparent transparency of the universe to $\gamma$-rays may possibly be explained by photons oscillating into axions relatively close to the AGN, followed by the axions traveling unimpeded through the universe and subsequently converting back to photons in the galactic or intergalactic magnetic fields. Such a scenario is possible for $m_a \lesssim 10^{-9}\,$eV and $M \approx 10^{11}-10^{12}\,$GeV \cite{Meyer:2013pny, Ringwald63, Ringwald64, Ringwald65, Ringwald66}.

Independently, there are observational hints for non-standard energy losses in white dwarfs, which may plausibly be explained by axions with $m_a < \keV$ for certain values of the axion coupling to electrons \cite{Ringwald53, Ringwald54, Ringwald55}. It is reasonable to assume that the axion-photon coupling and the axion-electron coupling are suppressed by the same scale, but with an unknown and model-dependent relative coefficient. In concrete models this coefficient may range from $0.1$ to $10^{-4}$ in the favour of stronger coupling to photons. Thus, the white dwarf hint may cautiously be interpreted as a hint for axions with  $m_a<\keV$  and $M\sim 10^{11}-10^{13}\,$GeV.

In the following, we will consider the restrictions from the laboratory bound $M>10^{10}\,$GeV together with the supernova $\gamma$-burst bound of $M>10^{11}\,$GeV. We will find that the CAB explanation for the soft excess is possible for light axions with $M \sim 10^{11}-7\ti10^{12}\,$GeV, which is in the same range as suggested by  the white dwarf energy loss hint and the $\gamma$-ray transparency hint.

Furthermore, while in our model axion-photon conversion in the Coma magnetic field should explain the soft X-ray excess, strong restrictions on the model parameters can be obtained by noting that higher energy photons should not be abundantly produced from axion-photon conversion: the excess is soft and does not survive to higher energies. This poses a restriction on the support of the CAB spectrum, as parametrised by the mean CAB energy. Here, we will impose that axion-photon conversion in the $0.5 - 0.6\keV$ band should not contribute to more than $10\%$ of the thermal luminosity in this range. More accurate --- and quite possibly more stringent --- bounds may be obtained by detailed extraction of the soft excess based on dedicated templates for the CAB spectrum.

In figure \ref{fig:MvsE}, we show the values for $M$ necessary to normalise the total CAB-converted luminosity to the observed soft excess for a wide range of mean axion energies and for all three models of the magnetic field.  In all cases we have considered $\Delta N_{\rm eff} = 0.5$, and for all models the parameters $\eta = 0.7$, $B_0 = 5.4\,\mug$ have been chosen.

The astrophysical and laboratory bounds on $M$ may then be translated into a lower bound on the mean CAB energy, and we find that, quite model-independently, this gives \hbox{$\langle E_{\rm CAB} \rangle \gtrsim 45-50\,$eV.} Meanwhile, the bound from X-ray overproduction gives more model-dependent constraints, but allows for \hbox{$\langle E_{\rm CAB} \rangle \lesssim 250\,$eV} for the interesting Model 3. We note that this restricts the values of $\lambda$ of equation (\ref{eq:scale}) to the range $2 \ti 10^5 < \lambda< 10^6$.

 In sum, we may then express the interesting values for the scale $M$ as
\beq
  10^{11}\,{\rm GeV} \lesssim M \lesssim  7 \ti 10^{12}  \sqrt{\frac{\Delta N^{(a)}_{\rm eff}}{0.5}}\,{\rm GeV} \, . \label{eq:boundM}
\eeq
Here we emphasise that $\Delta N^{(a)}_{\rm eff}$ denotes the extra relativistic contribution to the energy density from the axions responsible for the soft excess. If multiple species contribute to the dark radiation of the universe, then clearly $\Delta N_{\rm eff}^{(a)} < \Delta N_{\rm eff}$.

\begin{figure}[t]
\centering
\includegraphics[width=0.6\textwidth]{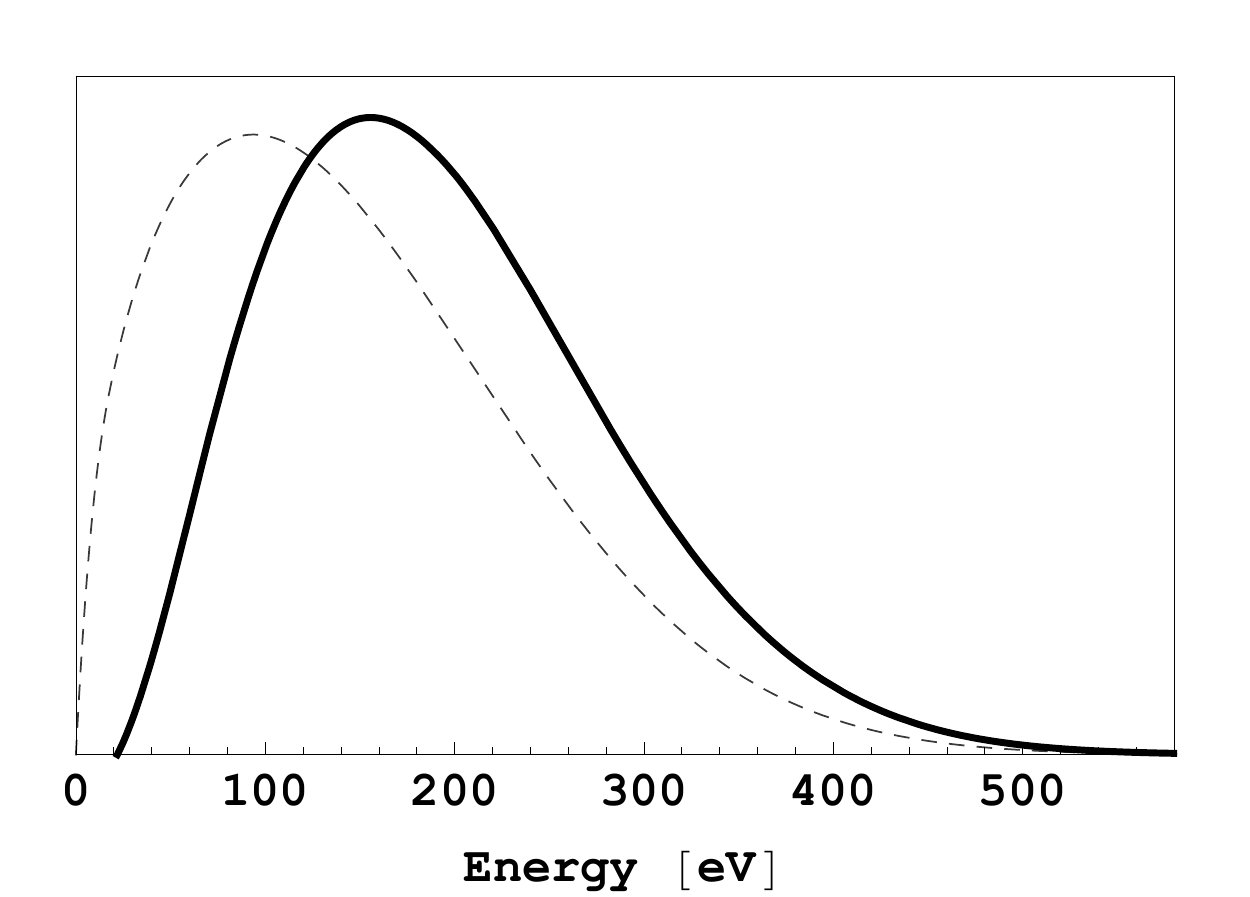}
\caption{The shape of the soft X-ray spectrum from axion-photon conversion in galaxy clusters (solid line), together with the ambient CAB (dashed)  for $\langle E_{\rm CAB} \rangle = 150\,$eV as obtained from propagation through the Model 3 magnetic field. Both curves have been normalised independently. }
\label{fig:axionphotonPDFs}
\end{figure}

Finally, let us comment on the spectral distribution of the soft excess photons, as predicted by axion-photon conversion of a CAB. Figure \ref{fig:axionphotonPDFs} shows the differential densities of the CAB as well as the photon spectrum arising after axion-photon conversion in the cluster magnetic field. We note that as low-energy axions have smaller conversion probabilities than high-energy axions, the resulting photon spectrum appears shifted to higher energies. For the particular case of   $\langle E_{\rm CAB} \rangle = 150\,$eV, the photon distribution has negligible support at $\omega < 21$ eV. In this case the functional form of the resulting photon distribution function is to a very good approximation given by
\beq
\frac{\dd n_{\gamma}}{\dd \omega} \sim \left(\frac{\omega-21\,{\rm eV}}{150\,{\rm eV}}\right)^{1.21} \exp\left[ -\left( \frac{\omega-21\,{\rm eV}}{170\,{\rm eV}}\right)^{1.78} \right]
\eeq
for $\omega \geq 21$ eV. The corresponding mean photon energy is $\langle E_{\gamma} \rangle \approx 200$ eV.

%%%%%%%%%%%%%%%%%%%%%%%%%%%%%%%%%%%%%%%%%%%%%%%%%%%%%%%%%%%

%%%%%%%%%%%%%%%%%%%%%%%%%%%%%%%%%%%%%%%%%%%%%%%%%%%%%%%%%%%
%%%%%%%%%%%%%%%%%%%%%%%%%%%%%%%%%%%%%%%%%%%%%%%%%%%%%%%%%%%

\subsection{Summary of results}
\label{sec:implications}
Let us conclude this section by summarising our results. We have found that the success of the CAB explanation for the morphology of the cluster soft excess depends on  some of the details of the cluster magnetic field, and in particular on the distribution of the coherence lengths of the magnetic domains.
The overall luminosity of the soft excess can easily be reproduced.
However, the morphology obtained from axion-photon conversion is compatible with the observed soft excess for magnetic field models which predominantly have short (transverse) coherence lengths of a few kpc, as well as models which have uniformly distributed (transverse) coherence lengths from a few to $100\kpc$.  For Gaussian magnetic field models with transverse coherence lengths predominantly in the $10\kpc$ range, such as our Model 1 above, the CAB explanation does not provide a close match to the
observed soft excess morphology in Coma.

Interestingly, the range of axion-photon couplings required to explain the cluster soft excess (c.f~equation (\ref{eq:boundM}))  is compatible with those suggested by the anomalously fast white dwarf cooling and with the anomalous transparency of the universe in $\gamma$-rays.

%Let us conclude this section by summarising our results. We have found that the viability of the CAB explanation for the morphology of the cluster soft excess depends on  some of the details of the cluster magnetic field, and in particular on the distribution of the coherence lengths of the magnetic domains. The luminosity obtained from axion-photon conversion is compatible with the observed soft excess for magnetic fields models which predominantly have short coherence lengths of a few kpc, as well as models which have uniformly distributed coherence lengths from a few to $100$ kpc.  Gaussian magnetic field models with coherence lengths predominantly in $10$ kpc range, such as our Model 1 above,  the CAB explanation does not  provide a close match the observed soft excess in Coma. 

%It is possible that more realistic, non-Gaussian models of the cluster magnetic field (such as those obtained from magnetohydrodynamic simulations), could make the CAB explanation viable also in this case. 

%The range of axion-photon couplings required to explain the cluster soft excess (c.f~equation (\ref{eq:boundM}))  is compatible with those suggested by the anomalously fast white dwarf cooling and with the anomalous transparency of the universe in $\gamma$-rays.

\section{Conclusions} \label{sec:concl}

In this paper, we have considered the possibility that axion-photon conversion of a string-theory-motivated Cosmic Axion Background with $0.1 - 1\keV$ energies may explain the long-standing soft X-ray excess in galaxy clusters.

We have focused in detail on the well-studied Coma cluster, for which the soft excess has been established at high
statistical significance, and for which rather elaborate stochastic magnetic field models have been constructed and have been shown to be consistent with observations of
Faraday rotation measures. Using these magnetic field models, we have
propagated axions through the cluster and studied the resulting morphology of soft X-ray photons.

This study has led us to three main conclusions. First, we have  confirmed the assertion of \cite{CAB} that the overall
luminosity of the soft excess can easily be explained by axion-photon conversion. For example, a 
CAB with mean energy of $\langle E_{\rm CAB} \rangle \approx 150\,$eV may explain the soft excess for
an axion-photon coupling of
$M \sim \sqrt{\frac{\Delta N_{\rm eff}}{0.5}}\, 6 \ti 10^{12}\,\hbox{GeV}$.

Second, the CAB-induced soft excess exhibits a non-trivial morphology which is sensitive to the distribution of coherence lengths of the transverse part of the
cluster magnetic field. Within the class of
Gaussian magnetic field models that are equally consistent with Faraday rotation measures \cite{Murgia:2004, 10020594}, those with a flat distribution in $k$-space were shown to reproduce well the observed soft excess morphology. On the contrary, for a turbulent Kolmogorov spectrum for the magnetic field (which locates power predominantly on larger, ${\cal O}(10\kpc)$, scales), the simulated excess flux disagrees with the observed morphology. 
The axionic explanation of the soft excess then requires either a flatter power spectrum or a shorter coherence length for the transverse component of the magnetic field (note that 
Faraday rotation measures constrain only the magnetic field component parallel to the line of sight).

Third,
the requirement that the cluster soft excess originates from a CAB strongly constrains the CAB properties. The absence of an X-ray excess at $E \gtrsim 0.5\keV$,
and the astrophysical requirement that the axion-photon coupling satisfy $M \gtrsim 10^{11}\,\hbox{GeV}$, constrains the mean CAB energy to the range
\bea
50\,  \hbox{eV} \lesssim \langle E_{\rm CAB} \rangle \lesssim 250\,\hbox{eV} \, ,
\eea
so that the CAB explanation of the soft excess is viable for
$10^{11}\,{\rm GeV} \lesssim M \lesssim  7 \ti 10^{12}  \sqrt{\frac{\Delta N^{(a)}_{\rm eff}}{0.5}}\,{\rm GeV}$.
The axion mass is similarly constrained to $m_a \lesssim 10^{-12}\,\hbox{eV}$.
Since the CAB spectral shape and the resulting soft X-ray spectrum are theoretically well-determined, the CAB hypothesis constitutes a sharp, well-defined, and predictive model that is subject to observational tests.

There are many further ways to extend this work and perform these tests. These include
\begin{itemize}
\item
Studies of axion propagation through other clusters for which the magnetic field
profile can be determined using Faraday rotation measures and soft excess properties have been determined. In this paper we have determined the values
of $M$ and $\langle E_{\rm CAB} \rangle$ that are necessary to reproduce the Coma soft excess. We can apply these
 to other clusters in the sample of \cite{astroph0205473} to see whether, given the magnetic field model, these values of $M$ and
 $\langle E_{\rm CAB} \rangle$ predict the correct magnitude for the soft excess.
\item
The magnetic field profile used here is by construction Gaussian and does not arise from a magnetohydrodynamical (MHD) simulation.
 It would be interesting to take magnetic field values found from
a numerical MHD simulation of cluster formation and see whether there are any qualitative differences in the resulting axion-photon conversion probabilities,
in particular in the behaviour of parallel and transverse coherence lengths.
\item
If a CAB exists, it continuously passes through our galaxy and converts to photons in the galactic magnetic field.
It would be interesting to use models of the galactic magnetic field
(for example those in \cite{12043662, 12107820}) to determine the resulting number and distribution of soft X-ray photons, given the $M$ and $\langle E_{\rm CAB} \rangle$
values required for the Coma cluster. This could then be compared
with counts from the ROSAT $0.25\keV$ all-sky survey. Preliminary work in this direction has been performed in \cite{13104464}.
\item
The sample of \cite{astroph0205473} contains 38 clusters for which soft excess studies have been performed. If information can be obtained about the
magnetic field and electron density in these clusters, it would be interesting to see whether this can be correlated with the
presence or absence of a soft excess.
\end{itemize}

The above studies will determine whether the CAB properties that can generate the soft excess in the Coma cluster remain consistent when applied to other
observations.

\section*{Acknowledgments}

We would like to thank the many people who have enhanced our knowledge of astrophysics, the soft excess, magnetic fields and galaxy clusters.
We are grateful for discussions with Pedro Alvarez, Armen Atoyan, Daniel Baumann, Katherine Blundell,
Stu Bowyer, Michele Cicoli, Ryan Cooke, Francesca Day, Jo Dunkley, Babette D\"obrich, Malcolm Fairbairn, Pedro Ferreira, Mark Goodsell, Arthur Hebecker, David Kraljic, Liam McAllister, Enrico Pajer, Stefano Profumo, Fernando Quevedo, Philip Rooney, Markus Rummel, Martin Sahl\'en, Alex Schekochihin, Katrien Steenbrugge, TJ Torres, Henry Tye, Scott Watson, Fabio Zandanel and Konstantin Zioutas. We would also like to thank Jonathan Patterson greatly for his help in running of the magnetic field code.

JC is funded by a Royal Society University Research Fellowship.
JC, DM, and LW are funded by the European Research Council starting grant `Supersymmetry Breaking in String Theory'.
SA and AP are funded by STFC studentships. Contents
reflect only the authors' views and not the views of the European Commission.

%%%%%%%%%%%%%%%%%%%%%%%%%%%%%%%%%%%%%%%%%%%%%%%%%%%%%%%%%%%
%%%%%%%%%%%%%%%%%%%%%%%%%%%%%%%%%%%%%%%%%%%%%%%%%%%%%%%%%%%
%%%%%%%%%%%%%%%%%%%%%%%%%%%%%%%%%%%%%%%%%%%%%%%%%%%%%%%%%%%

\bibliographystyle{JHEP}
\bibliography{CAB_LW_1212}

\end{document}